\documentclass{aptpub}
    \usepackage[T1]{fontenc}
    \usepackage{color}
    \usepackage{amsfonts}
    \usepackage{graphicx}
    \usepackage{latexsym,color}

    \authornames{
     }
    \shorttitle{ Testing for dependence in stable innovations processes }

    \numberwithin{equation}{section}  

    \numberwithin{figure}{section}
    
\begin{document}
    \begin{center}
    \title{Testing second order dynamics for autoregressive processes in presence of time-varying variance}
    \end{center} 
    \vspace*{0.1cm}
    \begin{center}
    Valentin Patilea$^{a}$ and Hamdi Ra\"{i}ssi$^{b}$\footnote{Corresponding author: 20, avenue des buttes de Co\"{e}smes, CS 70839, F-35708 Rennes Cedex 7,
    France. Email: hamdi.raissi@insa-rennes.fr} 
    \end{center}

    \begin{center}
    $^a$ CREST (Ensai) \& IRMAR (UEB)\\
    $^b$ IRMAR-INSA
    \end{center}

    \qquad

    \begin{center}
    \today
    \end{center}

    \qquad

    \begin{abstract}
    The volatility modeling for autoregressive univariate time series is considered.
    A benchmark approach is the stationary ARCH model of Engle (1982).
    Motivated by real data evidence, processes with non constant unconditional variance and ARCH effects have been recently introduced. 
    We take into account such possible non stationarity and propose simple testing procedures for ARCH effects.
    Adaptive McLeod and Li's portmanteau and ARCH-LM tests for checking for second order dynamics are provided. The standard versions of these tests, commonly used by practitioners, suppose constant unconditional variance. We  prove the failure of these standard tests  with time-varying unconditional variance. The theoretical results are illustrated by mean of simulated and real data.\\
    \end{abstract}

    \keywords{Adaptive tests ; Bandwidth selection; ARCH-LM tests; Kernel smoothing; Portmanteau tests; Time-varying variance.} 

    \quad

    \section{Introduction}
    \label{S2i}

    In the analysis of stationary time series nonlinearities are often taken into account by considering
    models which display conditionally heteroscedastic innovations. Reference can be made to the GARCH models introduced by Engle (1982) and Bollerslev (1986) or the
    All-Pass models studied by Andrews, Davis and Breidt (2006).
    Nevertheless using such models
    induces several important consequences for the time series analysis as pointed out in
    Francq, Roy and Zako\"{\i}an (2005) or Raïssi (2010a) among many others.
    Therefore tests for second order dynamics are needed to determine whether the whole dynamics of a stationary time series is captured by  modeling the variations in the level series, or nonlinear models have to be adjusted to the error process.

    It is well known that the squared residuals of stationary error processes with nonlinear effects are generally correlated.
    Two benchmark tests for testing the presence of second order dynamics are based on this simple remark: the Lagrange Multiplier test for detecting ARCH effects (ARCH-LM) proposed by Engle (1982) and the portmanteau test developed by McLeod and Li (1983) to detect correlation in the squared residuals. It is important to note that testing for second order dynamics in the error process is not equivalent to testing its independence.

    The benchmark ARCH-LM and portmanteau tests suppose the stationarity of the error process, an assumption that could be unrealistic in many cases.
    Numerous recent works have emphasized that time series often exhibit time-varying
    unconditional volatility. For instance Warnock and Warnock (2000) noted a reduction in volatility for U.S. employment data. Blanchard and Simon (2001) documented a substantial decline for U.S. output volatility.
    Sensier and van Dijk (2004) found that most of the 214 U.S. macroeconomic time series they considered have a break in volatility. Therefore tools taking into account such situations have been recently proposed. Kim and Park (2010) investigated cointegrated systems assuming smooth changes for the unconditional variance. The statistical inference of stock returns processes with possible time varying unconditional variance has been investigated by Engle and Rangel (2008), Mikosch and St\u{a}ric\u{a} (2004), St\u{a}ric\u{a} and Granger (2005) or Spokoiny (2009) among others. In particular St\u{a}ric\u{a} (2003) used a deterministic structure for the non-constant unconditional variance of stock returns and found that such a specification may be preferred to the usual stationary GARCH(1,1) for forecasting purposes. Chandler and Polonik (2012) proposed a test for investigating the modes of non constant unconditional variances.
    Francq and Gautier (2004) considered ARMA processes allowing for unconditional time-varying variance driven by a Markov chain. Berkes, Horv\'{a}th and Kokoszka (2004) proposed a test for detecting abrupt changes in the dynamics of the unconditional variance in a GARCH context. More precisely they tested the null hypothesis of standard stationary GARCH model against parameter change at a given date under the alternative which entails non constant unconditional variance. In the multivariate context Aue, H\"{o}rmann, Horv\`{a}th and Reimherr (2009) introduced a test procedure for break detection in the covariance structure.

    In the time series literature several contributions considered models with deterministic variance specification for the innovations to take into account the non constant unconditional variance.
    For instance Bai (2000) or Qu and Perron (2007) studied vector autoregressive processes with deterministic abrupt variance shifts. Xu and Phillips (2008) proposed adaptive  estimators for autoregressive parameters of stable univariate processes using kernel smoothing of the unconditional variance of the innovations. Raïssi (2010b) and Patilea and Raïssi (2011,2012) extended the work of Xu and Phillips and proposed modified tools for the usual specification-estimation-validation modeling cycle of multivariate autoregressive stable processes in the non standard framework of a nonparametric unconditional variance of the innovations. In these contributions the error processes are not allowed to exhibit second order dynamics and in general the asymptotic results break down in the presence of such nonlinear effects. Therefore using the tools developed assuming a deterministic volatility while second order dynamics are also present have no theoretical basis and could be misleading.
    On the other hand, models
    with stochastic
    volatility specification have been proposed.
    Engle and Rangel (2008) and Hafner and Linton (2010) among others studied models which display both non constant conditional and unconditional variance. Dahlhaus and Subba Rao (2006) investigated the properties of time varying ARCH (tvARCH) processes which are locally stationary but have a long run time varying unconditional variance change.

    In this paper we propose simple statistical tools allowing to analyze the volatility specification of a univariate time-series and decide between the two situations:  \emph{time-varying variance with or without second order (stochastic) dynamics}. These tools are also effective for detecting second order dynamics with constant unconditional variance. Our approaches are based on the following simple fact:
    stochastic volatility generally displays correlated squared residuals while no such correlations occur with deterministic time-varying volatility. Hence, new tests for second order dynamics are proposed as extensions of the benchmark Engle's ARCH-LM test and McLeod and Li's portmanteau test considered in the stationary case, and implemented in most specialized software. More precisely, we extend the ARCH-LM test to the case of possibly non stationary but stable autoregressive processes where the constant term in the ARCH structure is allowed to depend on time. Moreover, a modified portmanteau test based on the squared residuals which take into account non constant volatility is also developed. These new tests rely on the nonparametric kernel estimation of the unconditional volatility function. We prove that they are asymptotically chi-square distributed under the null hypothesis of no second second order dynamics. Our theoretical results are derived \emph{uniformly in the bandwidth} and hence we provide theoretical basis for data-driven bandwidths rules.
    The technical conditions imposed on the unconditional volatility are very mild and allow for general volatility patterns such like breaks, trends and cycles.
     It is also shown in this paper that in general the standard McLeod and Li  test statistic and the ARCH-LM test statistic tend to infinity as fast as  the series length under the null hypothesis of no second order dynamics, provided that the innovations variance depends on time. Therefore the standard tools for testing the presence of second order dynamics in the innovation process are very likely to spuriously reject the null hypothesis and should be avoided in the non standard, but quite realistic, framework considered herein.

    The structure of the paper is as follows. In section \ref{S2} the (unconditional) heteroscedastic autoregressive
    model is outlined. The unconditional variance specification of the errors process is discussed.
    In section \ref{S3} we extend the model by  allowing for an additional ARCH structure. This general model will be the framework for our
    ARCH-LM test, while the model with (unconditional) heteroscedasticity introduced in section  \ref{S2} will represent the null hypothesis.
    The portmanteau test we propose is introduced in the same framework though for this type of test the alternative hypothesis could be more general.
    For a more clear presentation, in section \ref{S3} we introduce our tests under the simplifying  assumption of known volatility structure. These tests with modified statistics that take into account the non constant unconditional volatility have standard asymptotic chi-square distributions. In section \ref{S4} the infeasible
    statistics based on known volatility are approximated using
    a kernel smoothing estimator of the volatility. Using these results,
    feasible adaptive tests for second order dynamics in the series are proposed.
    In section \ref{S5} we formally prove that in general the standard ARCH-LM  and McLeod and Li tests do not
    distinguish between unconditional heteroscedasticity with no second order
    dynamics effects and the case where the squared residuals are correlated.
    The finite sample properties of the new tests are investigated by mean of
    Monte Carlo experiments in section \ref{S6}. Moreover, the unreliability of
    the classical tests is illustrated. Some practical guidelines for a suitable
    choice of the bandwidth for the adaptive tests are given. Monte Carlo and
    bootstrap procedures improving the size of the adaptive tests are proposed.
    It turns out that the adaptive tests combined with the numerical methods for size correction are
    able to correctly take into account for time-varying unconditional variance when the second order dynamics are investigated.
    Moreover, the adaptive tests behave well when the unconditional variance is constant and thus the time series is stationary.
    Applications to several
    U.S. economic and financial real data sets are also presented: the second order
    dynamics of the innovations of the M1 monetary aggregate, the Producer Price
    Index (PPI) for all commodities and the Consumers
    Price Index (CPI) for all urban consumers for communication are studied. The technical proofs are gathered in the Appendix.


    \section{A time varying specification of the volatility}
    \label{S2}

    Let the univariate autoregressive process $(x_t)$ satisfy
    \begin{eqnarray}\label{AR}
    &&{x}_t={a}_{01}{x}_{t-1}+\dots+{a}_{0p}{x}_{t-p}+u_t\\&&
    u_t=h_t\epsilon_t,\nonumber
    \end{eqnarray}
    where ${a}_{0i}\in\mathbb{R}$, $1\leq i\leq p$, are
    such that $\det ({a}(z))\neq 0$ for all $|z|\leq 1$, with
    ${a}(z)=1-\sum_{i=1}^{p}{a}_{0i} z^i$. Note that all the results stated in the paper may be extended easily to the unit root case. We suppose that $x_{-p+1},\dots,x_1,\dots,x_n$ are observed.
    We also define $\mathcal{F}_{t}=\sigma(\epsilon_s,s\leq t)$ as the
    $\sigma$-field generated by $\{\epsilon_s,s\leq t\}$.
    Consider the following conditions on the innovations process $(u_t)$ where the rescaling approach of Dahlhaus (1997) is used for specifying the structure of the variance.\\

    \textbf{Assumption A1:} \;\; The $h_{t}$'s are given by $h_{t}=g(t/n)$,
    where $g(\cdot)$ is a
    measurable deter\-ministic and strictly positive function on the interval $(0,1]$, such that
    $\sup_{r\in(0,1]}|g(r)|<\infty$, and $g(\cdot)$ satisfies a piecewise Lipschitz
    condition on  $(0,1]$.\footnote{The piecewise Lipschitz condition means: there exists a positive integer $p$ and some mutually disjoint intervals $I_1,\dots,I_p$ with $I_1\cup\dots\cup I_p=(0,1]$ such that $g(r)=\sum_{l=1}^p g_l(r){\bf 1}_{\{r\in I_l\}},$ $r\in(0,1],$ where $g_1(\cdot),\dots,g_p(\cdot)$ are Lipschitz smooth functions on $I_1,\dots,I_p,$ respectively.} The process $(\epsilon_t)$ is assumed independent identically distributed (iid) of unit variance and such that $E(| \epsilon_t|^s)<\infty$ for some $s> 8$.\\

    Assumption {\bf A1} allows for a constant function $g(\cdot)$ which corresponds to the standard case of a stationary process $(x_t)$. In the sequel we will use $\epsilon$ to denote a generic random variable distributed as $\epsilon_t$. Under the Assumption {\bf A1} we have Cov$(u_t^2,u_{t-i}^2)=0$ for all $i\neq0$, and hence our assumption delineate the framework of the null hypothesis of our tests. A wide range of unconditional non constant variance is taken into account as for instance abrupt shifts, periodic or monotonic heteroscedasticity. Such framework was considered by
    Xu and Phillips (2008). Note that their assumptions $E(\epsilon_t|\mathcal{F}_{t-1})=0$ and
    $E(\epsilon_t^2|\mathcal{F}_{t-1})=1$ are more general but fairly close to our iid condition.
    Pesaran and Timmerman (2004) considered innovations with abrupt unconditional
    changes of the unconditional variance.
    Many applied papers assume piecewise constant volatility as in Batbekh, Osborn, Sensier and van Dijk (2007).
    Finally note that the autoregressive order $p$ of  model (\ref{AR}) is usually unknown. A well specified autoregressive order is important for testing for second order dynamics. However, the order $p$ can be identified and checked under {\bf A1} using the tools proposed in Raïssi (2010b) and
    Patilea and Raïssi (2011) and hence we will assume in the sequel that the lag length is well specified.

    In the stationary case it is well known that considering stochastic volatility for the error process entails second order dynamics in general (see e.g. Amendola and Francq (2009)). In the framework of non constant unconditional variance we note that the stochastic volatility specifications in Engle and Rangel (2008) or Boswijk and Zu (2007) entail
    Cov$(u_t^2,u_{t-i}^2)\neq0$ for some $i\neq0$.
    Hafner and Linton (2010) studied the following model
    \begin{equation}\label{complicated}
    u_t=\Sigma(t/n)^{1/2}G_t^{1/2}\epsilon_t,
    \end{equation}
    where $\Sigma(t/n)$ verifies similar conditions to $g(t/n)$ and $G_t$ follow a BEKK model (introduced by Engle and Kroner (1995)) allowing for second order dependence.

    Autoregressive processes with possibly second order dynamics in the innovations  could arise in many cases. It is well known that an invertible ARMA process can be approximated by
    an AR model with a large enough lag length. However assuming independent innovations for this AR model may appear too strong. Similarly, the assumption of independent errors is questionable in the important cases where an AR model is adjusted to a temporal aggregated time series. A further example arises when the univariate series is a component of a multivariate system with an error vector with cross-correlated components. Therefore it may be advisable to test for second order dynamics effects for the innovations in such situations.
    Considering model (\ref{AR}) to study the dynamics of a stable process with stochastic effects in the non constant unconditional variance innovations could be unreliable in many cases. Indeed some technical arguments used when assumptions like {\bf A1} are considered are no longer valid if  Cov$(u_t^2,u_{t-i}^2)\neq0$.
    In the stationary case it is well known that ignoring the presence of nonlinearities
    in the data can be quite misleading as pointed out by Francq and Raïssi (2007). It
    seems reasonable to imagine that a similar claim remains true with a time-varying variance. On
    the other hand considering a stochastic volatility as in (\ref{complicated}) may lead to
    unnecessary sophisticated structure and more complicated procedures, when second order dynamics are actually not present in the data.
    Finally, let us point out that in terms of level prediction model (\ref{AR}) imply that the best predictor for $x_{n+1}$ is linear, while it is well known that this is not always the case when nonlinearities are present in the data.
    In addition following St\u{a}ric\u{a} (2003) one can propose forecasts of the variance using the specification given by {\bf A1} which can be different by nature from the ones induced by some model allowing for stochastic effects for the variance as in Hafner and Linton (2010) section 7.1.
    The predictions delivered by modeling the variance structure have important applications in finance, as for instance for portfolio allocation or Value-at-Risk evaluation. As a consequence these different forecasting interpretations or methodologies could have implications for these tasks.

    In view of all the arguments enumerated above, we aim to provide simple tools which can help the
    practitioner to choose between a model specification only based on
    Assumption {\bf A1} and a specification which allow for second order
    dynamics together with unconditional heteroscedasticity.

    \section{Tests with the prior knowledge of the volatility structure}
    \label{S3}

    To better explain the differences with the stationary framework, in this section we suppose  that the true unconditional variance $h_t^2$ specified as in Assumption {\bf A1} is given. In the following section we show how $h_t^2$ could be estimated from data without changing the asymptotic behavior of the test statistics.

    Consider the  model\footnote{Like in the stationary case, see section 8 of Engle (1982), the model (\ref{ARlm1}) could be specified in a slightly more general way: $\tilde{h}_t^2 = h_t^2+ v(\alpha_{01}u_{t-1}^2+\dots+\alpha_{0m}u_{t-m}^2)$ where $v$ is some given positive differentiable function. For simplicity, we consider that $v(\cdot)$ is the identity function.}
    \begin{eqnarray}\label{ARlm1}
    &&{x}_t={a}_{01}{x}_{t-1}+\dots+{a}_{0p}{x}_{t-p}+u_t\\&&
    u_t=\tilde{h}_t\epsilon_t,\label{ARlm2}\\&&
    \tilde{h}_t^2=h_t^2+\alpha_{01}u_{t-1}^2+\dots+\alpha_{0m}u_{t-m}^2,\label{ARlm3}
    \end{eqnarray}
    where the $\alpha_{0i}$'s are assumed nonnegative and $h_t$ is defined as in Assumption {\bf A1}. Based on specification (\ref{ARlm3})
    stochastic effects in the volatility of the errors are tested  by considering the following pair of hypotheses
    \begin{equation}\label{eq_hyp}
    H_0: \alpha_{0i}=0,\quad \forall 1\leq i\leq m \qquad \mbox{vs.}\qquad H_1:\:\exists 1\leq i\leq m \text{ such that } \:\alpha_{0i}>0,
    \end{equation}
    for a given $m>0$ fixed by the practitioner.
    Under the null hypothesis we have $\tilde{h}_t=h_t$, that is  the non constant innovations variance with no second order dynamics described in the previous section.

    \subsection{LM-type test}

    First we consider the extension of the Lagrange Multiplier (LM) test proposed by Engle (1982) for detecting ARCH effects in stationary processes. In the classical ARCH-LM test the $h_t$'s are assumed constant under the conditional homoscedasticity hypothesis. On the other hand if we suppose that there exists $1\leq i\leq m$ such that $\alpha_{0i}>0$ and $g(\cdot)$ continuous in {\bf A1},
    the process $(u_t)$ generated by equations (\ref{ARlm2}) and (\ref{ARlm3}) is a particular case
    of the ARCH processes with time varying coefficients (tvARCH)
    studied by Dahlhaus and Subba Rao (2006).
    If we assume that the $h_t$'s have an abrupt change, we obtain a process which is closely related to
    the  ARCH($\infty$) structure with a change-point for the constant introduced in Kokoszka and Leipus (2000).
    If $\alpha_{0i}>0$ for some $i\in\{1,\dots,m\}$, the process $(u_t^2)$ is serially correlated. However it is important to recall that with the
    LM approach one does not estimate the full model with stochastic volatility given by (\ref{ARlm1})-(\ref{ARlm3}).

    Let $\theta_0$ and $\vartheta_0$ denote the true unknown values of
    the parameters $$\theta=(a_{1},\dots,a_{p})'\in\mathbb{R}^p \;\;
    \text{  and }\;\;
    \vartheta=(\alpha_{1},\dots,\alpha_{m})'\in\mathbb{R}^m.$$
    Let the gaussian quasi log-likelihood function (up to
    a constant) of the model (\ref{ARlm1})-(\ref{ARlm3})
    \begin{equation}\label{likeals}
    \mathcal{L}(\varphi)=-\frac{1}{2}\sum_{t=1}^nl_t\quad\mbox{with}\:l_t=
    \frac{u_t^2(\theta)}{\tilde{h}_t^2(\varphi)}+\log\left(\tilde{h}_t^2(\varphi)\right),
    \end{equation}
    for any $\varphi=(\theta',\vartheta')'$  and $$u_t(\theta)=x_t-a_1x_{t-1}-\dots-a_px_{t-p},\;\; \;\tilde{h}_t^2(\varphi)=h_t^2+\alpha_{1}u_{t-1}^2(\theta)+\dots+\alpha_{m}u_{t-m}^2(\theta),$$ provided  $\mathcal{L}(\varphi)$ exists.
    By definition, $u_t(\theta)=0$  when $t\leq0$. The (normalized) score vector for any constrained $\varphi^c=(\theta',0)'$ is given by
    \begin{eqnarray*}
    \mathcal{S}(\theta):=\frac{1}{\sqrt{n}}\frac{\partial\mathcal{L}(\varphi)}{\partial\vartheta}_{| \varphi=\varphi^c}&=&-\frac{1}{2\sqrt{n}}
    \sum_{t=1}^n\left\{-\frac{u_t^2(\theta)}{\tilde{h}_t^4(\varphi)}\frac{\partial \tilde{h}_t^2(\varphi)}{\partial \vartheta}+\frac{1}{\tilde{h}_t^2(\varphi)}\frac{\partial\tilde{h}_t^2(\varphi)}{\partial \vartheta}\right\}_{| \varphi=\varphi^c}\nonumber\\&=&\frac{1}{2\sqrt{n}}
    \sum_{t=1}^n\left(\frac{u_t^2(\theta)}{h_t^2}-1\right)U_{t-1}(\theta),
    \end{eqnarray*}
    where $U_{t-1}(\theta)=(u_{t-1}^2(\theta)/h_t^2,\dots,u_{t-m}^2(\theta)/h_t^2)'$. Let us introduce the generalized least-squares (GLS hereafter) estimator
    $$\hat{\theta}=\left\{\sum_{t=1}^nh_t^{-2}x_t\underline{x}_{t-1}\right\}
    \left\{\sum_{t=1}^nh_t^{-2}\underline{x}_{t-1}\underline{x}_{t-1}'\right\}^{-1},$$
    where $\underline{x}_t=(x_t,\dots,x_{t-p+1})$.
    Xu and Phillips (2008) showed that under Assumption {\bf A1}, and if $H_0$ is true, the GLS estimator is consistent and asymptotically normal with rate $\sqrt{n}$. In the sequel convergence in law is denoted by $\Rightarrow$.
    The proof of the following proposition is given in the Appendix.

    \begin{prop}\label{propostu} Under Assumption {\bf A1} and if the hypothesis $H_0$ holds true,
    $\mathcal{S}(\hat{\theta})\Rightarrow\mathcal{N}(0,\Sigma),$ where the asymptotic $m\times m$ covariance matrix is given by
    \begin{equation*}
    \Sigma=\frac{Var(\epsilon^2)}{4}\left(
                                                                            \begin{array}{cccc}
                                                                              E(\epsilon^4) & 1 & \cdots  &  1 \\
                                                                              1 & E(\epsilon^4) & \cdots  & 1 \\
                                                                              \vdots & \vdots & \ddots   &\vdots\\

                                                                              1 & 1 & \cdots &  E(\epsilon^4) \\
                                                                            \end{array}
                                                                          \right).
    \end{equation*}
    \end{prop}

    It is easy to see from (\ref{blues}) that Var$(\epsilon_t^2)$
    and $E(\epsilon_t^4)$ can be consistently estimated by
    $$\widehat{\mbox{Var}}(\epsilon^2)\! =\frac{1}{n}\sum_{t=1}^n\frac{u_t^4(\hat{\theta})}{h_t^4}-1,
    \:\mbox{ or alternatively }\:\:\widehat{\mbox{Var}}(\epsilon^2)=\frac{1}{n}\sum_{t=1}^n\frac{u_t^4(\hat{\theta})}{h_t^4}-\!
    \left(\frac{1}{n}\sum_{t=1}^n\frac{u_t^2(\hat{\theta})}{h_t^2}\right)^{\!\!2}$$
    and
    $$\widehat{E}(\epsilon^4):=\frac{1}{n}\sum_{t=1}^n\frac{u_t^4(\hat{\theta})}{h_t^4},$$
    under the null hypothesis. Let $\hat{\Sigma}$ denote a consistent estimator of $\Sigma$. Since  $E(\epsilon^4)>E^2(\epsilon^2)=1,$  $\hat{\Sigma}$ is non singular at least asymptotically.

    Now we have all the elements required to introduce the LM test for second order dynamics in the innovations process with known variance structure. Consider the (infeasible) test statistic
    $$Q_{GLS}=S(\hat{\theta})'\hat{\Sigma}^{-1}S(\hat{\theta}).$$

    \begin{prop} Under the conditions of Proposition \ref{propostu}, $Q_{GLS}$ is asymptotically distributed as a $\chi_m^2$ random variable.
    \end{prop}

    The GLS LM test we propose
    consists to reject the null hypothesis of no second order dynamics for the innovations at the asymptotic level $\upsilon$ if
    $$P(\chi_m^2>Q_{GLS}\mid x_{-p+1},\cdots,x_1,\cdots,x_n)<\upsilon.$$

    Now let us consider
    the alternative hypothesis of second order dynamics for the process $(u_t)$. For ease of exposition we suppose that $u_t$ is observed and we consider the case $m=1$ with $\alpha_{01}>0$. Noting that $$u_t^2=h_t^2+\alpha_{01}u_{t-1}^2+\tilde{h}_t^2(\epsilon_t^2-1),$$
    we have

    \begin{equation}\label{andrews}
    \frac{1}{n}\sum_{t=1}^n\left(\frac{u_{t}^2}{h_t^2}-1\right)\frac{u_{t-1}^2}{h_t^2}
    =\frac{\alpha_{01}}{n}\sum_{t=1}^n\left(\frac{u_{t-1}}{h_t}\right)^4
    +\frac{1}{n}\sum_{t=1}^n\frac{(\epsilon_t^2-1)\tilde{h}_t^2u_{t-1}^2}{h_t^4}.
    \end{equation}
    We can also write
    \begin{equation}\label{descendre}
    u_t^2=\sum_{i=0}^{\infty}\alpha_{01}^ih_{t-i}^2\epsilon_{t-i}^2\dots\epsilon_t^2,
    \end{equation}
    taking $h_l$ constant for $l\leq0$ and provided that this sum exists. Let us introduce $\overline{\tau}^2=\sup_th_t^2$ and $\underline{\tau}^2=\inf_th_t^2$, which are such that $0<\underline{\tau}<\overline{\tau}<\infty$ by {\bf A1}. We have
    $$\frac{\alpha_{01}}{n\overline{\tau}^4}\sum_{t=1}^nu_{t-1}^4<
    \frac{\alpha_{01}}{n}\sum_{t=1}^n\left(\frac{u_{t-1}}{h_t}\right)^4<
    \frac{\alpha_{01}}{n\underline{\tau}^4}\sum_{t=1}^nu_{t-1}^4.$$
    Define $\overline{u}_t^2=\overline{\tau}^2\sum_{i=0}^{\infty}\alpha_{01}^i\epsilon_{t-i}^2\dots\epsilon_t^2$
    and $\underline{u}_t^2=\underline{\tau}^2
    \sum_{i=0}^{\infty}\alpha_{01}^i\epsilon_{t-i}^2\dots\epsilon_t^2$, so that
    \begin{equation}\label{encadrement}
    \frac{\alpha_{01}}{n\overline{\tau}^4}\sum_{t=1}^n\underline{u}_{t-1}^4<
    \frac{\alpha_{01}}{n}\sum_{t=1}^n\left(\frac{u_{t-1}}{h_t}\right)^4<
    \frac{\alpha_{01}}{n\underline{\tau}^4}\sum_{t=1}^n\overline{u}_{t-1}^4.
    \end{equation}
    Note that the processes $(\overline{u}_t)$ and $(\underline{u}_t)$ corresponds to stationary ARCH(1) processes with autoregressive parameter $\alpha_{01}$. Therefore by the ergodic theorem the upper and the lower bounds in (\ref{encadrement}) converge to some strictly positive constants, provided that
    $E(\mid \underline{u}_t\mid^{4r})<E(\mid \overline{u}_t\mid^{4r})<\infty$ for some $r>1$\footnote{This induces restrictions on the parameter $\alpha_{01}$. For instance we must have $\alpha_{01}<0.57$ if the process $(\epsilon_t)$ is standard Gaussian. If such restriction hold it is clear that the sum in (\ref{descendre})
    is well defined. However note that we do not require that
    $g(.)$ is continuous under the alternative.}. In particular it follows that

    $$\frac{\alpha_{01}}{n}\sum_{t=1}^n\left(\frac{u_{t-1}}{h_t}\right)^4\geq c+o_p(1),$$
    for some $c>0$. In the same way it can be shown that the weight matrix $\hat{\Sigma}$, defined using $\widehat{\mbox{Var}}(\epsilon^2)$ and
    $\widehat{E}(\epsilon^4)$, is bounded in probability. It is also clear that
    the second term on the right-hand side of (\ref{andrews}) is $o_p(1)$ since $(\epsilon_t^2-1)$ is independent of $\tilde{h}_t^2u_{t-1}^2$. Indeed we have $\tilde{h}_t^2=h_t^2+\sum_{i=1}^{\infty}\alpha_{01}^ih_{t-i}^2
    \epsilon_{t-1}^2\dots\epsilon_{t-i}^2$.
    Hence we have
    $Q_{GLS}\geq Cn+o_p(n)$ for some constant $C>0$ and this ensures the consistency of the GLS LM test.


    \subsection{McLeod and Li type test}

    In this part we extend the test of McLeod and Li (1983) to detect the possible presence of second order
    dynamics in the series. Define
    \begin{equation*}
    \bar{\omega}_4=\frac{n^{-1}\sum_{t=1}^nu_t^4(\hat{\theta})}{\widehat{E}(\epsilon^4)}=\int_0^1g^4(r)dr+o_p(1)
    \end{equation*}
    and
    \begin{equation*}
    \bar{\omega}_8=\frac{n^{-1}\sum_{t=1}^nu_t^8(\hat{\theta})}{\widehat{E}(\epsilon^8)}=\int_0^1g^8(r) dr+o_p(1)
    \end{equation*}
    with $\widehat{E}(\epsilon^8)=n^{-1}\sum_{t=1}^n u_t^8(\hat{\theta})h_t^{-8}$. The two $o_p(1)$ rates above could be obtained under {\bf A1}, and assuming that $H_0$ is true, using similar arguments to that in Lemma 1 and 2 of  Phillips and Xu (2006).

    Consider the following infeasible Ljung-Box type  portmanteau statistic
    $$Q_{GLS}^*=\left\{n(n+2)\sum_{i=1}^m\hat{r}^2(i)/(n-i)\right\}\frac{\bar{\omega}_4^2}{\bar{\omega}_8},$$
    where $\hat{r}(i)=\hat{\gamma}(i)/\hat{\gamma}(0)$ and
    \begin{equation*}
    \hat{\gamma}(j)=n^{-1}\sum_{t=1+j}^n\{u_t^2(\hat{\theta})-E(u_t^2)\}\{u_{t-j}^2(\hat{\theta})-E(u_{t-j}^2)\},\qquad 0\leq j\leq m,
    \end{equation*}
    for a given $m>0$ fixed by the practitioner. Let us recall that $E(u_t^2)=h_t^2$ under the null hypothesis. The following proposition gives the asymptotic critical values of the $Q_{GLS}^*$ statistic.

    \begin{prop}\label{propostwo}
    Under Assumption {\bf A1} and if $H_0$ holds true, $Q_{GLS}^*\Rightarrow\chi_m^2$.
    \end{prop}

    The asymptotic behavior of the $Q_{GLS}^*$ can be obtained straightforwardly considering a decomposition as in equation (\ref{blues}) in the Appendix and the following results which can be established using  arguments like in Lemma 1 and 2 of  Phillips and Xu (2006).

    \begin{lem} Under Assumption {\bf A1} and if $H_0$ holds true,
    \begin{equation*}
    \gamma_n(0):=n^{-1}\sum_{t=1}^n\{u_t^2-E(u_t^2)\}^2 = \{E(\epsilon^4)-1\} \int_0^1g^4(r)dr+o_p(1),
    \end{equation*}
    $$n^{\frac{1}{2}}\gamma_n(i):=n^{-\frac{1}{2}}
    \sum_{t=1}^n\{u_t^2-E(u_t^2)\}\{u_{t-i}^2-E(u_{t-i}^2)\}
    \Rightarrow\mathcal{N}\left(0,\sigma_\gamma^2\right)$$
    where
    $$\sigma_\gamma^2 = \{E(\epsilon^4)-1\}^2 \int_0^1 g^8(r)dr.$$
    In addition
    $$\lim_{n\to\infty}\mbox{Cov}(n^{\frac{1}{2}}\gamma_n(i),n^{\frac{1}{2}} \gamma_n(j))=0,\quad i\neq j.$$
    \end{lem}

    \quad

    Using Proposition \ref{propostwo}, one can propose a portmanteau test for second order dynamics based on the prior knowledge of $h_t$: reject the null hypothesis of no second order dynamics for the innovations at the asymptotic level $\upsilon$ if
    $$P(\chi_m^2>Q_{GLS}^*\mid x_{-p+1},\cdots,x_1,\cdots,x_n)<\upsilon.$$
    Under the alternative of second order dynamics for $(u_t)$ we also have $Q_{GLS}^*=C'n+o_p(n)$ for some $C'>0$ so that the GLS LB test is consistent. This can be seen by considering similar arguments to that used for the $Q_{GLS}$ statistic.

    \section{Adaptive tests for second order dynamics for time series with non constant variance}
    \label{S4}
    In this section we propose feasible tests based on approximations of the $Q_{GLS}$ and $Q_{GLS}^*$ statistics.
    We first provide an estimator of the unknown $h_t^2$ under $H_0$.
    Let us define the usual Ordinary Least Squares (OLS) estimator
    $$\check{\theta}=\left\{\sum_{t=1}^nx_t\underline{x}_{t-1}\right\}
    \left\{\sum_{t=1}^n\underline{x}_{t-1}\underline{x}_{t-1}'\right\}^{-1}$$
    and let $u_t(\check{\theta})$ be the OLS residuals. Following Xu and Phillips (2008) we define an
    adaptive estimator of the variance structure using the OLS residuals
    $$\hat{h}_t^2=\sum_{i=1}^nw_{ti}u_i^2(\check{\theta}),$$
    with $w_{ti}=\left(\sum_{j=1}^nK_{tj}\right)^{-1}K_{ti}$ and
    $$K_{ti}=\left\{
                  \begin{array}{c}
                    K((t-i)/nb)\quad \mbox{if}\quad t\neq i\\
                    0  \quad\mbox{if}\quad t=i,\\
                  \end{array}
                \right.$$
    where $K(\cdot)$ is a kernel function on the real line and $b$ is the bandwidth. Let us consider the adaptive estimator
    $$\tilde{\theta}=\left\{\sum_{t=1}^n\hat{h}_t^{-2}x_t\underline{x}_{t-1}\right\}
    \left\{\sum_{t=1}^n\hat{h}_t^{-2}\underline{x}_{t-1}
    \underline{x}_{t-1}'\right\}^{-1}.$$

    \vspace{0.3 cm}

    \textbf{Assumption A1':} Suppose that Assumption \textbf{A1} holds true
     and $$\inf_{r\in(0,1]} g(r)\geq c  >0$$ for some constant $c.$

    \vspace{0.3 cm}

    \textbf{Assumption A2:} \, (i) The kernel $K(\cdot)$ is a bounded density function defined on the real line such that $K(\cdot)$ is nondecreasing on $(-\infty, 0]$ and decreasing on $[0,\infty)$ and $\int_\mathbb{R} v^2K(v)dv < \infty$. The function $K(\cdot)$ is differentiable except a finite number of points and the derivative $K^\prime(\cdot)$  satisfies $\int_{\mathbb{R}}|x K^\prime (x)| dx < \infty.$
    Moreover, the Fourier Transform $\mathcal{F}[K](\cdot)$ of $K(\cdot)$ satisfies $\int_{\mathbb{R}}  \left| s\right|^\tau \left| \mathcal{F}[K](s) \right|ds <\infty$ for some $\tau>0$.

    (ii) The bandwidth $b$, $1\leq k\leq l\leq d$, are taken in the range $\mathfrak{B}_n = [c_{min} b_n, c_{max} b_n]$ with $0< c_{min}< c_{max}< \infty$ and $nb_n^{4-\gamma} + 1/nb_n^{2+\gamma} \rightarrow 0$ as $n\rightarrow \infty$, for some small $\gamma >0$.

    \vspace{0.3 cm}

    Under the Assumptions \textbf{A1'} and \textbf{A2}, Patilea and Raïssi (2011) showed that $$\sqrt{n}(\hat{\theta}-\tilde{\theta})=o_p(1),$$ \emph{uniformly} with respect to $b\in\mathfrak{B}_n$.\footnote{For proving that   $\tilde{\theta} - \hat \theta$ is negligible the condition  $nb_n^{4-\gamma}  \rightarrow 0$ could be replaced by $b_n \rightarrow 0$ while $K^\prime$ integrable may replace the condition $\int_{\mathbb{R}}|x K^\prime (x)|  dx < \infty.$
    The more restrictive conditions we impose here on the bandwidth and the kernel will serve in the proof of Proposition \ref{proposthree} for rendering the bias induced by the nonparametric smoothing negligible.} Most of the common kernels used in practice satisfy the technical conditions imposed in  Assumption \textbf{A2}-(i). The uniformity with respect to the bandwidth provides a theoretical basis for data-driven bandwidth rules, for instance for bandwidth obtained cross-validation as proposed in section \ref{S6} below.

    We are now able to introduce the adaptive tests for second order dynamics. Define the following adaptive ARCH-LM statistic which approximate the infeasible GLS ARCH-LM statistic
    $$Q_{ALS}=\mathcal{\widetilde{S}}(\tilde{\theta})'
    \widetilde{\Sigma}^{-1}\mathcal{\widetilde{S}}(\tilde{\theta}),$$
    where

    \begin{equation}\label{scoreadapt}
    \mathcal{\widetilde{S}}(\theta)=\frac{1}{2\sqrt{n}}
    \sum_{t=1}^n\left(\frac{u_t^2(\theta)}{\hat{h}_t^2}-1\right)
    \widetilde{U}_{t-1}(\theta)
    \end{equation}
    with
    $\widetilde{U}_{t-1}(\theta)=(u_{t-1}^2(\theta)/\hat{h}_t^2,\dots
    ,u_{t-m}^2(\theta)/\hat{h}_t^2)'$ for any $\theta\in\mathbb{R}^p$ is
    the approximated score vector. The weight matrix is given by

    $$\widetilde{\Sigma}=\frac{\widetilde{Var}(\epsilon^2)}{4}\left(
                                                                             \begin{array}{cccc}
                                                                              \widetilde E(\epsilon^4) & 1 & \cdots  &  1 \\
                                                                              1 & \widetilde E(\epsilon^4) & \cdots  & 1 \\
                                                                              \vdots & \vdots & \ddots   &\vdots\\

                                                                              1 & 1 & \cdots &  \widetilde E(\epsilon^4) \\
                                                                            \end{array}
                                                                          \right),$$
    where

    $$\widetilde{Var}(\epsilon^2)=n^{-1}\sum_{t=1}^n\frac{u_t^4(\tilde{\theta})}{\hat{h}_t^4}-
    \left(n^{-1}\sum_{t=1}^n\frac{u_t^2(\tilde{\theta})}{\hat{h}_t^2}\right)^2\quad\mbox{and}\quad
    \widetilde{E}(\epsilon^4)=n^{-1}\sum_{t=1}^n\frac{u_t^4(\tilde{\theta})}{\hat{h}_t^4}$$
    can be taken. Define
    \begin{equation}\label{eqqqe}
    \tilde{\omega}_4=\frac{n^{-1}\sum_{t=1}^nu_t^4(\tilde{\theta})}
    {\widetilde{E}(\epsilon^4)}=\int_0^1g^4(r)dr+o_p(1)
    \end{equation}
    and
    $$\tilde{\omega}_8=\frac{n^{-1}\sum_{t=1}^nu_t^8(\tilde{\theta})}
    {\widetilde{E}(\epsilon^8)}=\int_0^1g^8(r) dr+o_p(1)$$
    with $\widetilde{E}(\epsilon^8)=n^{-1}\sum_{t=1}^n
    u_t^8(\tilde{\theta})\hat{h}_t^{-8}$. We also consider the adaptive
    portmanteau test statistic which approximate the infeasible
    portmanteau statistic

    $$Q_{ALS}^*=\left\{n(n+2)\sum_{i=1}^m\tilde{r}^2(i)/(n-i)\right\}
    \frac{\tilde{\omega}_4^2}{\tilde{\omega}_8},$$ where as above
    $\tilde{r}(i)=\tilde{\gamma}(i)/\tilde{\gamma}(0)$ and
    \begin{equation}\label{autocovariances}
    \tilde{\gamma}(i)=\frac{1}{n}\sum_{t=1+i}^n(u_t^2(\tilde{\theta})-
    \widehat{E}(u_t^2))(u_{t-i}^2(\tilde{\theta})-\widehat{E}(u_{t-i}^2)),
    \end{equation}
    where $\widehat{E}(u_t^2)=\hat{h}_t^2$ can be taken. In the following proposition we state the asymptotic equivalence between the adaptive test statistics and the infeasible test statistics based on the knowledge of the variance structure. This equivalence is obtained \emph{uniformly} with respect to $b\in\mathfrak{B}_n.$

    \begin{prop}\label{proposthree}
    Under Assumptions \textbf{A1'} and \textbf{A2} and if $H_0$ holds true, for any fixed $m\geq 1$,  $Q_{ALS}=Q_{GLS}+o_p(1)$ and $Q_{ALS}^*=Q_{GLS}^*+o_p(1)$ uniformly with respect to $b\in\mathfrak{B}_n.$ Consequently,  $Q_{ALS}, Q_{ALS}^*\Rightarrow\chi_m^2$ for any data-driven sequence of bandwidths in $\mathfrak{B}_n.$
      \end{prop}

    Proposition \ref{proposthree} provides feasible versions of the ARCH-LM and portmanteau tests introduced in section \ref{S3} for testing for second order dependence in the presence of time-varying variance. The proof given in the Appendix remains valid in the standard case of a constant unconditional variance.

    \section{Inadequacy of the standard tests}
    \label{S5}

    The benchmark Engle's ARCH-LM test and McLeod and Li's portmanteau
    test are implemented in most specialized software and
    the standard statistics are compared with $\chi^2-$type critical
    values. In this section we show that when $h_t$ is not constant (for
    instance $h_t$ is piecewise constant with a single change-point), these two
    standard statistics grow to infinity as fast as $n$. This means that
    in the presence of non constant time-varying volatility the standard
    versions of ARCH-LM and McLeod and Li tests spuriously reject the
    null hypothesis of non second order dynamics effects with
    probability tending to 1. It is well-known that the standard ARCH-LM
    statistic is equivalent to a standard portmanteau statistic, see for
    instance section 5.4.2 of Francq and Zako\"{\i}an (2010). Hence it
    will suffice to investigate the failure of the classical McLeod and
    Li portmanteau statistic in the presence of time-varying variance,
    the same conclusion will apply  to the standard ARCH-LM statistic.
    The theoretical investigation presented in this section will be
    completed by empirical examples in the following section.

    Let $\bar{\omega}_2=n^{-1}\sum_{t=1}^n u_{t}^2(\check{\theta})=\int_0^1g^2(r)dr+o_p(1).$ Consider the standard portmanteau test
    $$Q_S^*=n(n+2)\sum_{i=1}^mr_S^2(i)/(n-i),$$
    where $r_S(i)=\gamma_S(i)/\gamma_S(0)$ and
    $$\gamma_S(i)=\frac{1}{n}\sum_{t=1+i}^n(u_t^2-\bar{\omega}_2)(u_{t-i}^2-\bar{\omega}_2),\quad i\in\{0,1,\dots,m\}.$$
    Let write
    \begin{eqnarray*}
    n^{\frac{1}{2}}\gamma_S (i)&=&
    n^{-\frac{1}{2}}\sum_{t=1+i}^nh_t^2h_{t-i}^2(\epsilon_t^2-1)(\epsilon_{t-i}^2-1)+
    n^{-\frac{1}{2}}\sum_{t=1+i}^n(h_t^2-\bar{\omega}_2)h_{t-i}^2(\epsilon_{t-i}^2-1)\\&+&
    n^{-\frac{1}{2}}\sum_{t=1+i}^n(h_{t-i}^2-\bar{\omega}_2)h_t^2(\epsilon_t^2-1)+
    n^{-\frac{1}{2}}\sum_{t=1+i}^n(h_t^2-\bar{\omega}_2)(h_{t-i}^2-\bar{\omega}_2)\\
    &=:& \Gamma_{1n}(i) + \cdots + \Gamma_{4n}(i), \qquad\mbox{say}.
    \end{eqnarray*}
    The term $\Gamma_{4n}(i)$  could be written
    $$
    \Gamma_{4n}(i) =  n^{-\frac{1}{2}}\!\!\sum_{t=1+i}^n\! \left(h_t^2-\int_0^1g^2(r)dr\!\right)\! \left(h_{t-i}^2-\int_0^1g^2(r)dr\!\right)  + o_p(1) =: \widetilde \Gamma_{4n}(i) + o_p (1)
    $$
    where $\widetilde \Gamma_{4n}(i)$ is deterministic. By arguments similar to those used for equation (\ref{eqqqe}), under {\bf A1} we have
    \begin{equation}\label{Michelle}
    n^{-\frac{1}{2}}\widetilde \Gamma_{4n}(i) = \int_0^1 g^4(r) dr - \left( \int_0^1 g^2(r) dr \right)^2 + o(1),
    \end{equation}
    for any $1\leq i\leq m$. Since $\int_0^1 g^4(r)dr - \left( \int_0^1 g^2 (r) dr \right)^2 >0$ if $h_t$ is not constant, deduce that for all
    $1\leq i \leq m,$
    $\Gamma_{4n}(i) = c n^{\frac{1}{2}}  + o_p(n^{\frac{1}{2}})$ with $c$ a strictly positive constant. Using the Lindeberg CLT and the Slutsky Lemma, deduce that for any $i$,  $ \Gamma_{1n}(i)$ to $\Gamma_{3n}(i)$ are asymptotically normally distributed with zero mean and some complicated asymptotic variances. It follows that $Q_S^* = C n + o_p(n)$ for some strictly positive constant $C.$

\section{Numerical illustrations}
\label{S6}

We conducted extensive empirical experiments using simulated and real data sets to study the performances and to illustrate the new test procedures introduced above. Two aspects were investigated using simulated samples: data-driven bandwidths choices and finite sample corrections of the tests level. The lessons we obtained from the simulations were applied to the US economic series studied hereinafter.

In practice, the adaptive procedures we propose require a bandwidth selection rule. There is a huge statistical literature on how to select the bandwidth when using kernel smoothing for regression functions.
However, there is no reason to expect that a bandwidth that is `optimal' for regression estimation purposes would automatically yield reasonable level and power in a testing procedure, in particular for the ones we consider herein. That is why we considered two types of data-driven bandwidth rules. First we followed  the classical cross-validation (CV) approach and we searched for the bandwidth  minimizing a CV criterion like $\sum_{t=1}^n| \hat{h}_t^2-u_t^2(\check{\theta})|^2.$
The approach would produce bandwidths $b$ that decrease as fast as $n^{-1/5}$, see H\"{a}rdle, Marron (1985). That is why we fixed the set $\mathfrak{B}_n$ defined in Assumption \textbf{A2} to be of the form $[c_{min}n^{-1/5}, \; c_{max}n^{-1/5}],$ for some constants $c_{min}, c_{max},$ and we minimized the CV criterion over this range.

Second we propose a kind of `rule of thumb' calibrated to provide accurate levels. Our rule of thumb searches a bandwidth like $b= \gamma (\hat \sigma^2/n)^{1/5}$ where $\hat \sigma^2 $ is the empirical variance of $\hat u_t^2,$ $t=1,...,n.$ The constant $\gamma$ belongs to some finite grid of positive numbers and is calibrated by simulation in order to obtain accurate levels. Clearly, there is no reason to expect that the same constant $\gamma$ will be selected for the two types of tests that we consider. In subsection \ref{S8} we explain how to perform this calibration with real data.

Concerning the finite sample corrections of the critical values, two approaches are proposed. On one hand, we considered  bootstrap counterparts for our adaptive tests. More precisely, we follow the simple procedure described in Francq and Zako\"{\i}an (2010, p335) in a stationary GARCH context for generating bootstrap data. In short the bootstrap statistics are computed according to the following steps:

\begin{itemize}
\item[1-] Compute $\hat{\epsilon}_t=u_t(\tilde{\theta})/\hat{h}_t$ for $t=1,\dots,n$.
\item[2-] Generate $\epsilon_t^{(b)}$'s for $t=1,\dots,n$, by drawing randomly with replacement from $\hat{\epsilon}_1,\dots,\hat{\epsilon}_n$.
\item[3-] Construct the bootstrap residuals $\hat{u}_t^{(b)}=\epsilon_t^{(b)}\hat{h}_t$ for $t=1,\dots,n$ and the bootstrap series of $x_t$ using the autoregressive model.
\item[4-] Build the kernel estimator $\hat{h}_t^{(b)}$'s from the OLS residuals obtained with bootstrap series. Next, compute the adaptive estimator $\tilde\theta^{(b)}.$

\item[5-] Compute the score vector $\widetilde{\mathcal{S}}^{(b)}$ (resp. the autocorrelations $\tilde{r}(i)^{(b)},$ $i=1,\cdots,m$) as in (\ref{scoreadapt}) (resp. as in (\ref{autocovariances})) using the $u_t(\tilde\theta^{(b)})$'s and the $\hat{h}_t^{(b)}$'s.
\item[6-]
Compute the bootstrap version $Q^{(b)}_{ALS}$ of the adaptive LM test statistics  (resp. the bootstrap version $Q^{*(b)}_{ALS}$ of the adaptive portmanteau test statistic).

\item[7-] Repeat the steps 2 to 6   $B$ times for some large $B$. Use $Q^{*}_{ALS}$ and $Q^{*(b)}_{ALS}$'s (resp. $Q_{ALS}$ and $Q^{(b)}_{ALS}$'s) to compute the bootstrap $p$-values of the portmanteau (resp. LM) test.
\end{itemize}
In our simulations we considered $B=499$ bootstrap iterations. To avoid estimating higher order moments that may introduce some instability in the results, we preferred the following versions of the adaptive test statistics
\begin{equation}\label{test_stat_mod}
\mathcal{Q}_{ALS}=\widetilde{\mathcal{S}}(\tilde \theta)^\prime
    \widetilde{\mathcal{S}}(\tilde \theta)\quad  \text{and } \quad \mathcal{Q}^{*}_{ALS}=n(n+2)\sum_{i=1}^m\tilde{r}^2(i)/(n-i)
\end{equation}
and their bootstrap counterparts. Under the assumptions of Proposition \ref{proposthree} we have
\begin{equation*}
\mathcal{Q}_{ALS}\Rightarrow\sum_{i=1}^m\kappa_iU_i^2
\quad\mbox{and}\quad
\mathcal{Q}_{ALS}^*\Rightarrow\delta\chi_m^2,
\end{equation*}
where $\delta=\int_0^1g^8(r)dr/\left(\int_0^1g^4(r)dr\right)^2$, the $U_i$'s are independent $\mathcal{N}(0,1)$ variables and the $\kappa_i$'s are the
eigenvalues of the matrix $\Sigma$ defined in Proposition \ref{propostu}.

On the other hand, we used the following Monte Carlo procedure to replicate the asymptotic distribution of the test statistics:
\begin{itemize}
\item[(a)]
Draw $\eta_t$, $t=1,\dots,n,$  iid centered random variables of variance 1 and finite moment of order 3; the $\eta_t$'s are independent of the observations; to generate these variables we used the distribution introduced by Mammen (1993), that is  $P(\eta_t=-(\sqrt{5}-1)/2)=(\sqrt{5}+1)/(2\sqrt{5})$ and
$P(\eta_t= (\sqrt{5}+1)/2)= (\sqrt{5}-1)/(2\sqrt{5}).$

\item[(b)]
Define $$\xi_t = \eta_t (u_t^2(\tilde{\theta})/\hat h_t^2-1)\widetilde{U}_{t-1}^*(\tilde{\theta})\;\; \text{ and } \;\; \zeta_t (i)=\eta_t\{u_t^2(\tilde{\theta})-\hat h_t^2\}
\eta_{t-i}\{u_{t-i}^2(\tilde{\theta})-\hat h_{t-i}^2\},$$
where $\widetilde{U}_{t-1}^*(\tilde{\theta})=
(\eta_{t-1}(u_{t-1}^2(\tilde{\theta})/\hat{h}_t^2-1)+1,...,
\eta_{t-m}(u_{t-m}^2(\tilde{\theta})/\hat{h}_t^2-1)+1)$.
Consider
$$\widetilde {\mathcal{S}}(\tilde{\theta})^{(r)}=\frac{1}{2\sqrt{n}}
\sum_{t=1}^n \xi_t\quad
\mbox{and}\quad\tilde{r}(i)^{(r)}=\tilde{\gamma}(i)^{(r)}/\tilde{\gamma}(0)^{(r)}$$ with
$\tilde{\gamma}(i)^{(r)}=n^{-1}\sum_{t=1+i}^n\zeta_t (i).$

\item[(c)] Compute the statistics $\mathcal{Q}^{(r)}_{ALS}$ and $\mathcal{Q}^{*(r)}_{ALS}$ like in (\ref{test_stat_mod}) but with $\widetilde {\mathcal{S}}(\tilde{\theta})$ and $\hat{r}(i)$ replaced by
    $\widetilde {\mathcal{S}}(\tilde{\theta})^{(r)}$ and $\hat{r}(i)^{(r)}$, respectively.
\item[(d)] Repeat steps (a) to (c) $R$ times for some large $R$. Use $\mathcal{Q}^{*}_{ALS}$ and $\mathcal{Q}^{*(r)}_{ALS}$'s (resp. $\mathcal{Q}_{ALS}$ and $\mathcal{Q}^{(r)}_{ALS}$'s) to compute the $p$-values of the portmanteau (resp. LM) test.
\end{itemize}
In our simulations we considered $R=499$ Monte Carlo iterations. The Monte Carlo procedure presented above does not require to resample the observations $x_t$'s, so that the possible problem of explosive processes with roots within the unit circle is avoided.
However,  the autoregressive order $p$ in (\ref{AR}) should be carefully fitted for the original data to avoid spurious rejections. Patilea and Ra\"{i}ssi (2011) and Ra\"{i}ssi (2010) provide suitable tools for testing linear restrictions on the autoregressive coefficients $a_{1},\dots,a_{p}$ and choosing the autoregressive order $p$ in the presence of time-varying variance.

In the sequel we denote by $LM_S$ and $LB_S$ the standard ARCH-LM and portmanteau tests. Similarly,
let $LM_{GLS},$  $LM_{ALS},$ $LB_{GLS}$ and $LB_{ALS}$ be the modified GLS and ALS tests we introduced in Sections \ref{S3} and \ref{S4}.
The results for the infeasible $LM_{GLS}$ and $LB_{GLS}$ tests represent a benchmark for our adaptive approach.
As pointed out above the ALS tests are subject to bandwidth choice and finite sample improvements. The subscripts "sm" and "la" are used for relatively small or large bandwidth choices of the form $b=\gamma (\hat \sigma^2/n)^{1/5}.$
We use the subscript "$cv$" when the bandwidth is selected by cross-validation. The superscripts "$b$" and "$mc$" denote the use of the bootstrap and Monte Carlo methods described above for the ALS tests.  In summary $LM_{cv,ALS}$, $LM_{cv,ALS}^{mc}$ and $LB_{cv,ALS}$, $LB_{cv,ALS}^{mc}$ correspond to tests for which the bandwidth is selected automatically from the data. Tests like $LM_{sm,ALS}^{b}$, $LM_{la,ALS}^{b}$, and $LB_{sm,ALS}^{mc}$, $LB_{la,ALS}^{mc}$ correspond to tests
with bandwidths $b=\gamma (\hat \sigma^2/n)^{1/5}$ for suitable $\gamma$. The constant $\gamma$
for such bandwidths may be fixed in practice using a preliminary simulation experiment as illustrated in the real data analysis below.\footnote{Tests based on bandwidth selected by cross-validation together with bootstrap critical values could be also considered. However, such tests require much more computation time in applications and hence will be omitted.}

\subsection{Simulation experiments}\label{S7}

To assess the finite sample performances of the tests considered in this paper, we simulate $N=1000$ independent trajectories of lengths $n=100$, $n=200$ and $n=500$ following the model
\begin{eqnarray*}
&&u_t=\tilde{h}_t\epsilon_t,\\&&
\tilde{h}_t^2=h_t^2+\alpha_{0}u_{t-1}^2,
\end{eqnarray*}
where the process $(\epsilon_t)$ is iid standard Gaussian. The simulated data $(u_t)$ have non constant unconditional variance if the
$h_t$'s change over time. Recall that the $h_t$'s are driven by some deterministic function $g(r)$ with $r\in\left[0,1\right]$. For the homoscedastic case we take $g(r)=20,$ $\forall r\in(0,1] $.
The time-varying variance specification has a sinusoidal form $g(r)=30-10\sin(1.5\pi r+\pi/6)(1+r)$. In view of numerous data sets we
investigated  (most of them not reported here) this variance specification produces trajectories that mimic some realistic features. In particular the sinusoidal form reproduces the variance clustering sometimes observed in real data. In all the experiments we test the null hypothesis of no second order dynamics for the innovations process at the (asymptotic) level 5\%. With this specification we retained the values
$\gamma=0.12$ (resp. $\gamma=0.2$) for determining small (resp. large) bandwidths $b$.

\subsubsection{The behavior under the null hypothesis\\}
\quad First,  we set
$\alpha_{0}=0$ and $\tilde{h}_t^2=h_t^2$. In such situation there is no second order
dynamics, that is Cov$(u_t^2,u_{t-i}^2)=0$ for all $i\neq0$.
The results are provided in Table \ref{tab2} for the heteroscedastic case, and in Table \ref{tab1} for the homoscedastic case.
Table \ref{tab2} indicates that the standard tests tend to reject spuriously the null hypothesis as the sample size is increased when the unconditional variance is not constant. This illustrates the results of Section \ref{S5} where we formally proved the inadequacy of the standard tests in our non standard framework.

Concerning the portmanteau tests, we note that the very simple $LB_{cv,ALS}$ test with automatic bandwidth cross-validation selection deliver quite good results for $m=1$. In general the $LB_{cv,ALS}$ and $LB_{GLS}$ have similar results in our non standard case. In particular for large $m,$ the $LB_{cv,ALS}$ and $LB_{GLS}$ tests become oversized. This size distortion can be corrected using the Monte Carlo approach for the adaptive tests.
When a rule of thumb is used for the bandwidth selection we found that small $\gamma$ is preferable to obtain relative rejection frequencies close to the 5\% for large $m$. Let us point out that the bandwidths corresponding to small $\gamma$ are generally close to the bandwidths selected by cross-validation.

On the other hand for the $LM$ tests, we remark that the tests with automatic bandwidth choice by cross-validation do not give satisfactory results.
It is found that the Monte Carlo corrections are disappointing in all the cases. On the other hand if the bandwidth $b=\gamma(\hat \sigma^2/n)^{1/5}$ is selected using some appropriate rule of thumb, it appears that the bootstrap corrections improve the results.
More precisely it turns out that the $\gamma$ have to be taken large to obtain relative rejections frequencies which are in general close to the 5\% level when the unconditional variance is time-varying.
It is interesting to note that in such a case large $\gamma$ give bandwidths which are somewhat far from the bandwidths selected by cross-validation.
In conclusion for the $LM$ tests it emerges that only bootstrap correction can provide adaptive tests sharing the nice properties of the $LM_{GLS}$ test.

Finally if the unconditional variance is in fact constant it turns out from Table \ref{tab1} that the adaptive $LM$ tests with bootstrap correction have generally satisfactory results when compared to the valid standard ARCH-LM test. We also remark that the adaptive and standard portmanteau tests have in general comparable results in the homoscedastic case. Therefore the size accuracy of the tests providing a good control of the error of first kind in the non standard case is not deteriorated when the unconditional variance is constant.\\

\subsubsection{The behavior under the alternative hypothesis\\}
\begin{figure}[h]\!\!\!\!\!\!\!\!\!\!
\hspace*{1.5cm}$m=1$\hspace*{2.7cm}$m=3$ \hspace*{2.7cm}$m=6$
\vspace*{8.7cm}

\protect \includegraphics{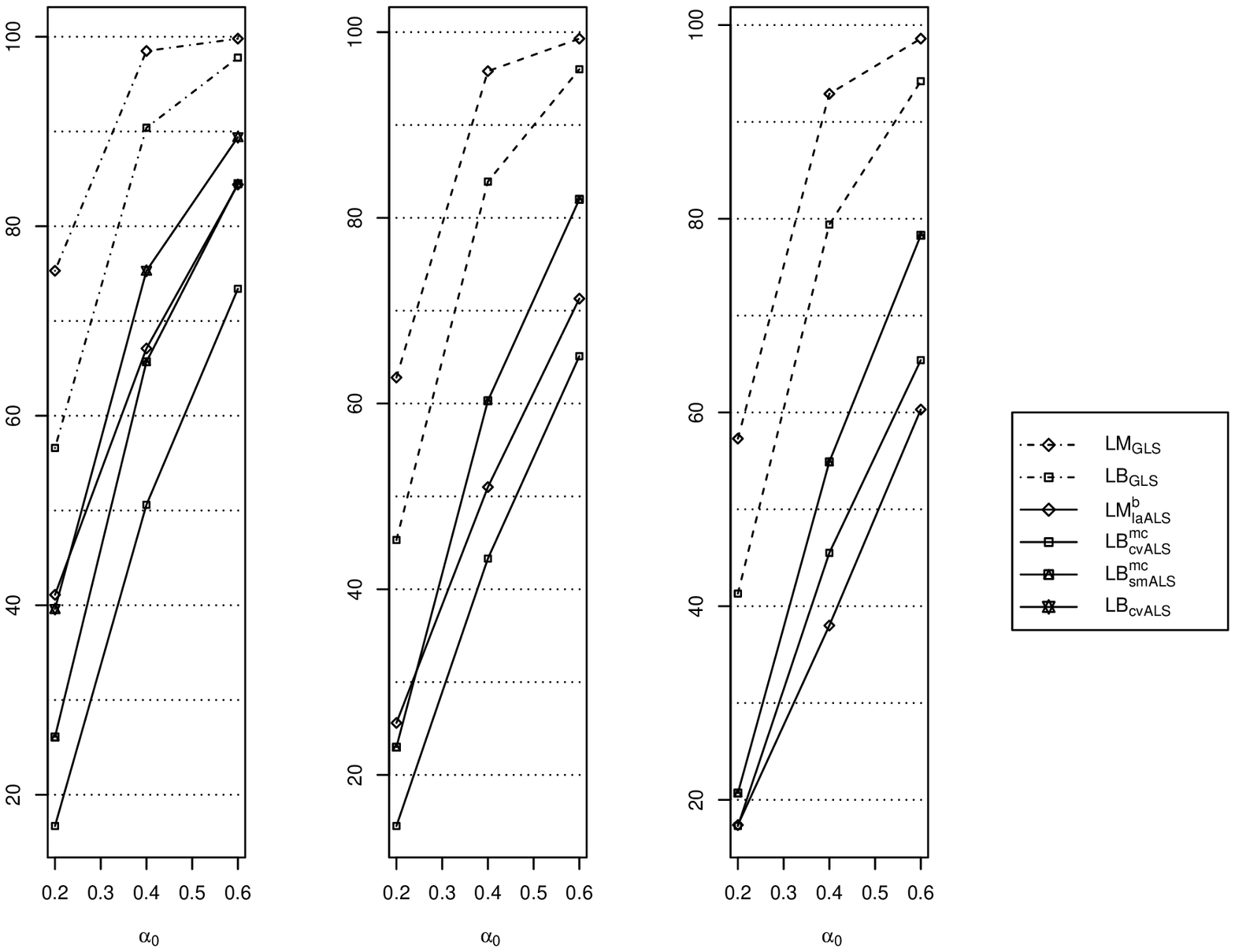}
\caption{\label{powerhetero}
{\footnotesize The empirical power of the tests for second order dynamics
in the case of non constant unconditional variance with $n=200$.
}}
\end{figure}

\begin{figure}[h]\!\!\!\!\!\!\!\!\!\!
\vspace*{0.3cm}\hspace*{1.5cm}$m=1$\hspace*{2.7cm}$m=3$ \hspace*{2.7cm}$m=6$
\vspace*{8.7cm}

\protect \includegraphics{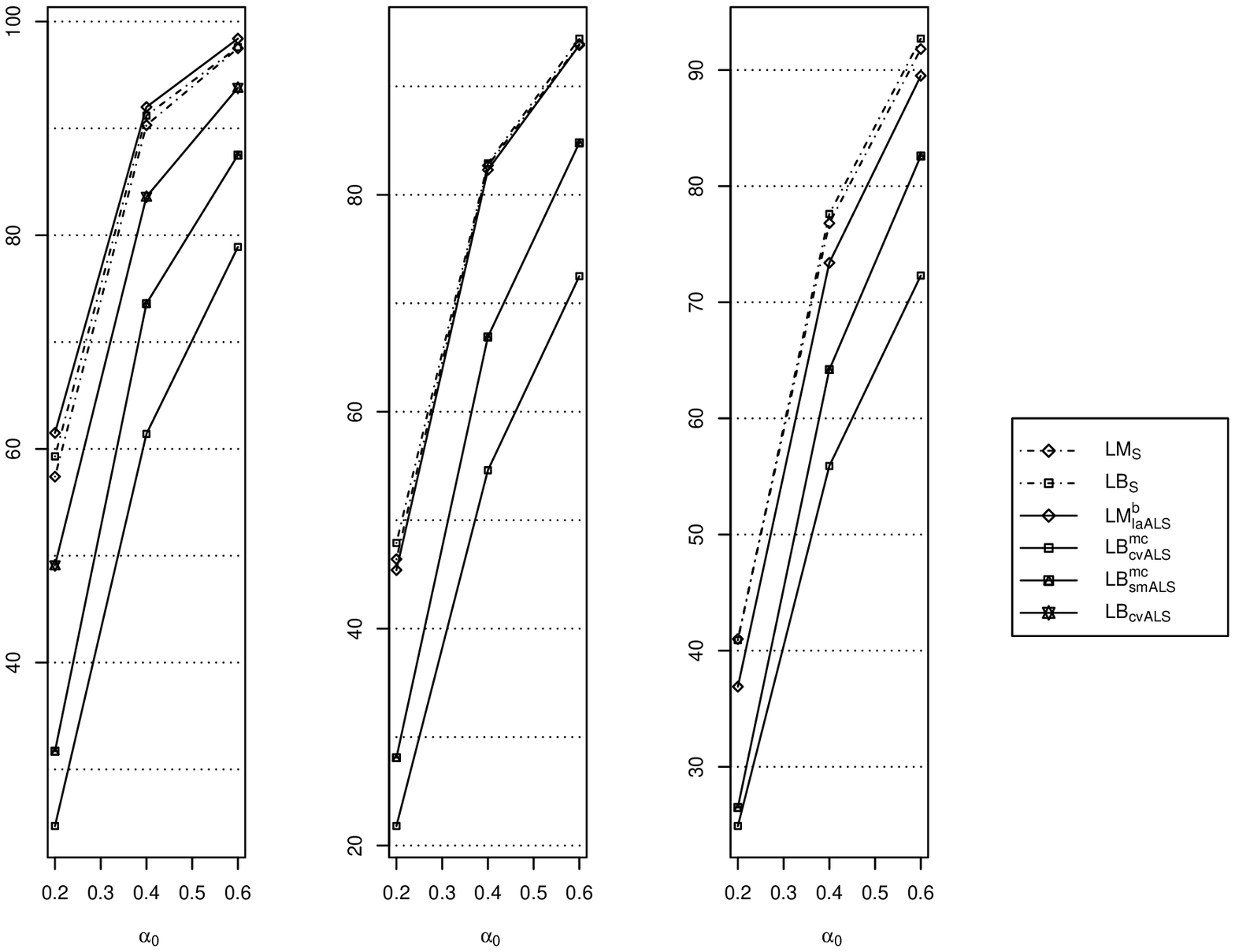}
\caption{\label{powerhomo}
{\footnotesize The empirical power of the tests for second order dynamics
in the case of constant unconditional variance with $n=200$.
}}
\end{figure}

\quad
Now we turn to experiments with second order
dynamics for the  process $(u_t)$ and we take
$\alpha_{0}=0.2,0.4,0.6$, with again $m=1,3,5$.
The variance is set time-varying and constant as above.
We focus on the power properties of the $LM_{la,ALS}^{b}$, $LB_{cv,ALS}^{mc}$ and $LB_{sm,ALS}^{mc}$ tests that showed a good control of the type I errors in almost all the studied situations. The outputs for the simple $LB_{cv,ALS}$ test for $m=1$ are also displayed.
The results for the infeasible GLS tests are given when the variance is not constant, while the
standard tests are considered for the homoscedastic case.
We only report the case with sample size $n=200,$ the results for $n=100$ and $n=500$ are similar.
The results are presented in Figure \ref{powerhetero} for the heteroscedastic case, and in Figure \ref{powerhomo} and for the homoscedastic case.

From Figure \ref{powerhetero} it appears that the GLS tests are more powerful than the ALS tests. Nevertheless recall that
the GLS tests are built on the unrealistic assumption of known $h_t$ which make them infeasible in applications.
Now considering the outputs for the homoscedastic case in Figure \ref{powerhomo}, it turns out that the standard tests
have some power advantage on most of the adaptive portmanteau tests.
However, we noted that the $LM_{la,ALS}^b$ has similar properties to the standard tests, which
may be explained by the fact that larger bandwidths allows for more accurate estimates
of a constant variance function.
Finally it seems that all the tests are subject to a loss of power as $m$ is increased.\\

In the light of our simulation experiments we could draw some conclusions and provide some guidelines for the real data applications.
The standard ARCH-LM and McLeod and Li's tests are not able to distinguish between deterministic non constant variance and the case where there exists additional second order dynamics in the series. Therefore, if there is some clear statistical evidence or an underlying information on the data showing that the unconditional variance is not constant, the alternative adaptive tests developed in this paper should replace the standard procedures.
Clearly, if the unconditional variance is constant the simple standard tests are preferable.
However, we did not found a major loss of performance for some of our more sophisticated tests in the standard case.
Therefore, in the case of a doubt on the constancy of the variance, again the adaptive tests should be preferred.

The adaptive tests we propose depend on a bandwidth and their performances are clearly influenced by the bandwidth choice. Our simulations show that the adaptive portmanteau tests behave well in general when the bandwidth is automatically selected by cross-validation ($m=1$). When $m$ is taken large Monte Carlo corrections for the $LB$ tests provide quite good results. These tests have the advantage that the bandwidth is selected automatically from the data. Portmanteau tests with bandwidth chosen by fixing some appropriate multiplicative constant can be used as well. However, we noted that in general the appropriate constant leads to a bandwidth close to the one obtained by cross-validation. Hence we recommend the simple bandwidth selection by cross-validation for the $LB$ approach. For the $LM$ tests the cross-validation bandwidth selection yields poor results. For the LM approach the bandwidth has to be fixed using some numerical
method to guarantee a good control of type I errors.
\\


\subsection{Real data applications}\label{S8}

Several applications of the tools developed in this
paper to real data are presented below.
The presence of second order
dynamics for series which obviously exhibit non constant unconditional variance is tested.
We discarded the adaptive tests which have disappointing results in the light of our simulation experiments. More precisely the results of
the $LB_{cv,ALS}^{mc}$ will be used for the tests with automatic bandwidth selection. The $LM_{la,ALS}^b$
or $LB_{sm,ALS}^{mc}$ are considered when the bandwidths are obtained by fixing
the multiplicative constant $\gamma$. The standard tests are
reported for comparison. The constant $\gamma$ is fixed following the data-driven procedure:
\begin{enumerate}
  \item[1)] Estimate the unconditional time-varying variance by
  minimizing the CV criterion.
  \item[2)] Approximate the deterministic variance function $g^2(\cdot)$ defined in {\bf A1} using  for instance the Lagrange interpolation of some points of the estimated variance. Let us denote by $\mathfrak{\hat{h}}_t^2$ the approximate variance.
  \item[3)] Compute  $\hat{\varepsilon}_t=u_t(\tilde{\theta})/\mathfrak{\hat{h}}_t$ for $t=1,\dots,n$ and use these values to generate processes $ \hat u_t^{(b)}$ with non constant variance many times following the steps  (2) and (3) in the bootstrap procedure described above.
  \item[4)] For each process generated in the previous step estimate the variance with several bandwidths corresponding to several values of $\gamma$ and consequently implement the LB or LM tests using the Monte-Carlo or bootstrap procedures.
  \item[5)] Compute the rejection frequencies for each $\gamma$. Select the value $\gamma$ for which the rejection frequency is close to the desired nominal level.
\end{enumerate}

For simplicity, in this data-driven procedure for calibrating a suitable value of $\gamma$ we focus on the $(u_t)$ process and we do not generate artificial samples of $(x_t)$ and re-estimate the parameter $\theta$. The uncertainty due to $\theta$ is expected to be negligible in our context. Moreover, in order to avoid the computation burden, when calibrating the value of $\gamma$ one could use a subseries provided that it contains sufficient information allowing to approximate reasonably well the unconditional variance.
It is interesting to note that the $\gamma$ values selected through this procedure are generally different for the orders $m=1,3,5$ in the test statistics. This is not surprising in view of the outputs of our simulation experiments.
For all the tests we fixed a nominal level of 5\%. For the
bootstrap and Monte-Carlo tests we take $B=499$. The $p$-values are displayed in bold
type when they are lower than 5\%.\\

\begin{figure}[h]\!\!\!\!\!\!\!\!\!\!
\vspace*{10cm}

\protect \includegraphics{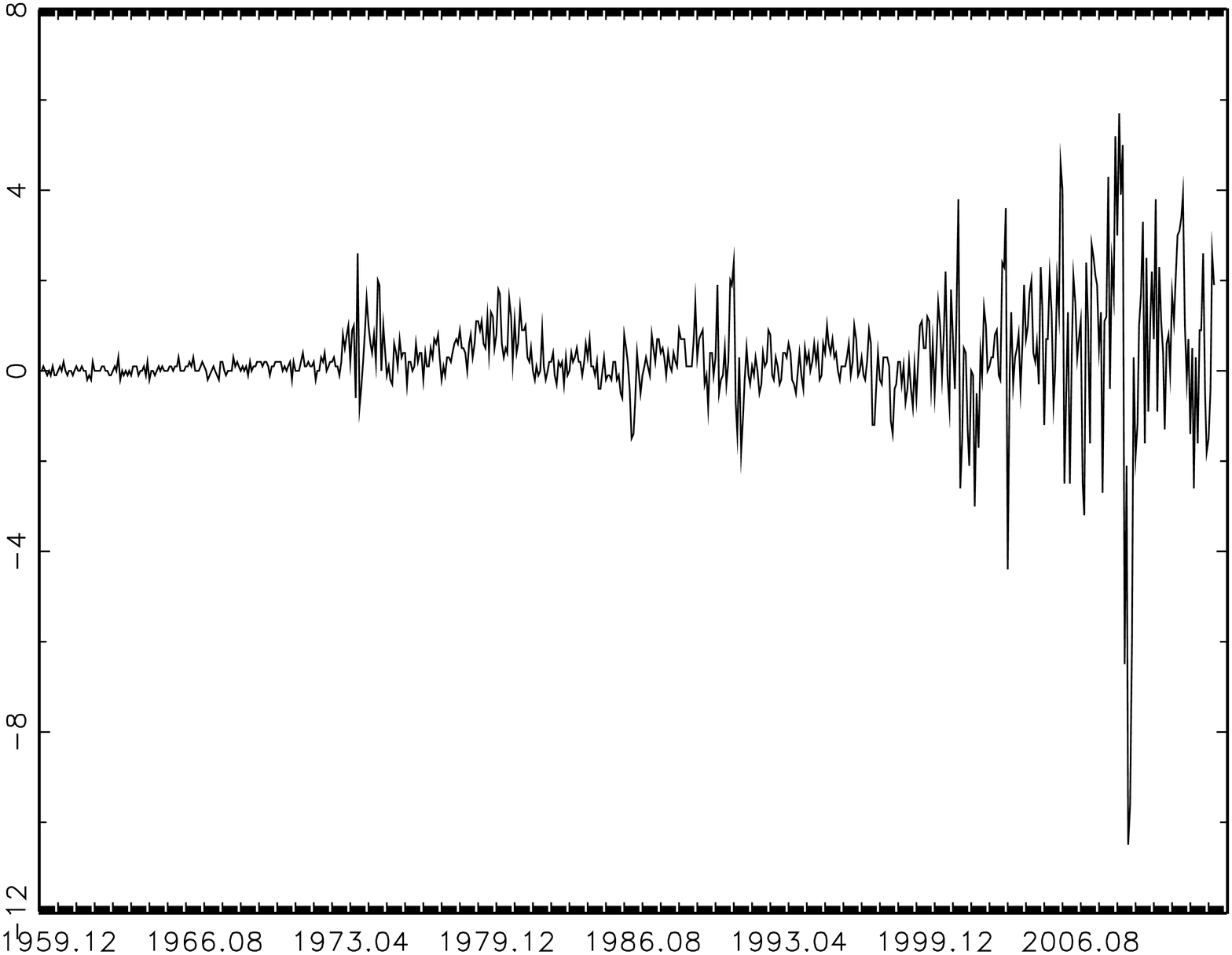} \protect \includegraphics{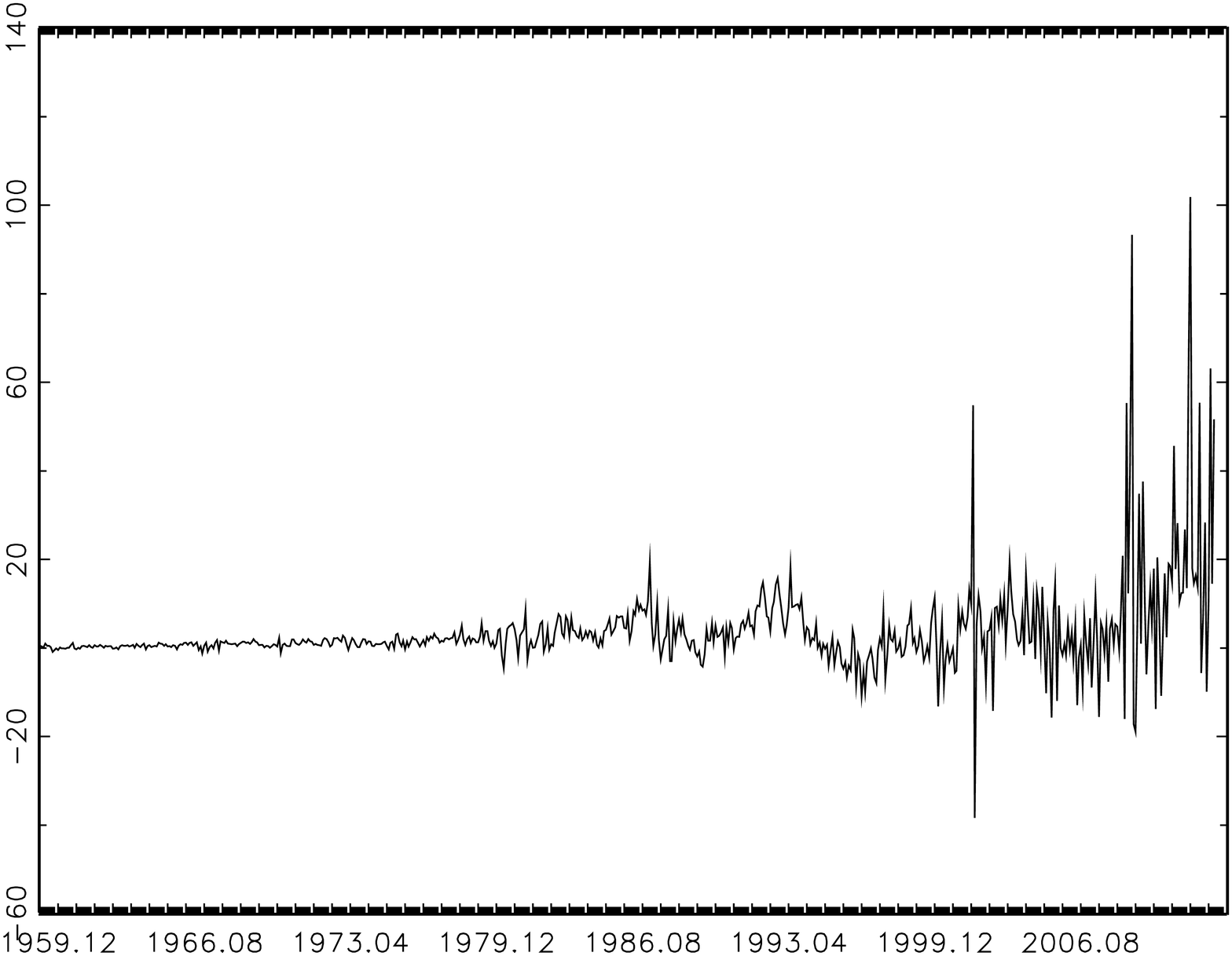} 
\protect \includegraphics{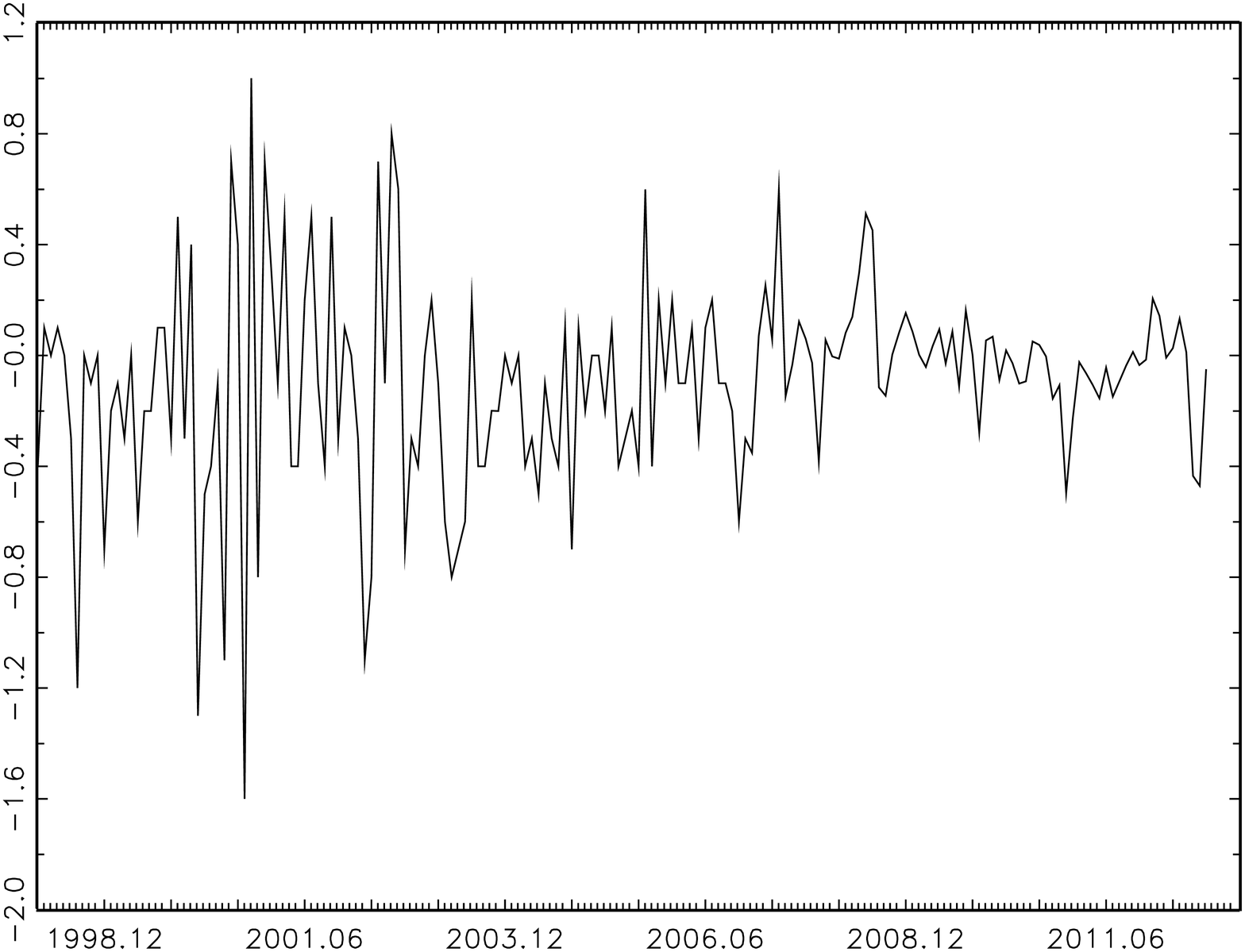} \caption{\label{data}
{\footnotesize The first differences of the studied data: The monthly PPI index on the top left panel and the monthly M1 on the top right panel from 2/1/1959 to 9/1/2012 ($n=644$). The monthly CPI from 2/1/1998 to 9/1/2012 is displayed on the bottom panel ($n=176$). Data source: The research division of the federal reserve bank of Saint Louis, www.research.stlouis.org.}}
\end{figure}

We investigated U.S. economic data: the first differences of the monthly M1 monetary aggregate and of the monthly Producer Price Index (the PPI for all commodities) from February 1, 1959 to September 1, 2012. The length of these series is $n=644$. The monthly Consumer Price Index (CPI) for all urban consumers for communication from February 1998 to September 2012 is also investigated. The length of the series is $n=176$ for the CPI data. Since the original CPI series has a random walk behavior, we considered
the series of the first differences for our analysis. The three series investigated are
plotted in Figure \ref{data}.\footnote{The data are available in the website of the research
division of the federal reserve bank of Saint Louis: www.research.stlouisfed.org, series ID: M1SL for the M1 and PPIACO for the PPI, and CUSR0000SAE2 for the CPI.}
Such series  are often studied
in the literature: see  L\"{u}tkepohl, Ter\"{a}svirta and Wolters (1999) and
references therein for the M1 data; see  Grier and Perry (2000) among others for the PPI series; the consumer price indexes
are included in many applied works.

In view of Figure \ref{data} it appears that the unconditional variance of the M1, PPI and and CPI is globally increasing.
From 1973 for the PPI and from 1979 for the M1, we also observe clusters of
large values which may indicate the presence of second order dynamics.
It also emerges that the CPI data display a declining variance. In order to study the variance structure of the innovations, we adjusted AR models to the PPI and M1 series. Meanwhile, it was found that the CPI series is uncorrelated. To check the AR models adequacy  we used the portmanteau tests developed in Patilea and Raïssi (2011) which are valid under {\bf A1} (the outputs are not displayed here).

Once the linear dynamics of the
series are well captured, we can turn to the analysis of the second order dynamics in the uncorrelated processes. For the variance specification of the series a practitioner would likely use a model which takes into account for possible non constant
unconditional variance but excludes second order dynamics as
in {\bf A1}. Nevertheless in view of the observed variance clustering for the PPI and M1 series, more elaborated
model specifications allowing for both conditional and
unconditional non constant variance could  be used, as in Engle and Rangel (2008) or Hafner and Linton (2010).
We aim to provide arguments for fitting an adequate model for
the variance by testing if second order dynamics are present in the innovations or not.

We applied the
$LB_{cv,ALS}^{mc}$, $LM_{la,ALS}^b$ and $LB_{sm,ALS}^{mc}$ tests
to the CPI series.
We also applied
the $LB_{cv,ALS}^{mc}$ test to the residuals of the PPI and M1 series.
The outputs of the adaptive tests are compared with those of the standard test: the results are given in Table \ref{pvaluestsod1} for the PPI and M1 and in Table \ref{pvaluestsod2} for the CPI. We first remark that the $p$-values of the standard tests are very close to zero so that the null hypothesis of no second order dynamics is clearly rejected for the three data series.
On the other hand, for the PPI and M1 data, for any of the values of $m$ we considered, the $p$-values of the adaptive tests are far above 0.05, so that the null hypothesis of no second order dynamics is not rejected. For the  CPI data, all but one $p$-values of the adaptive tests are above 0.05, some of them far above 0.05, the exception being the case $m=1$.

In view of Figure \ref{data} which clearly reveals a non stationary in variance behavior of the three series we investigated, it is likely that the standard tests spuriously reject the null hypothesis. Meanwhile, our adaptive tests provide very different conclusions which could lead one to confidently reconsider the possibility of using an elaborated specification which allow for second order dynamics for the M1 and PPI series. We obtain the same conclusion for the  CPI series, again on the contrary to the standard tests.

\newpage

    \section*{References}
    \begin{description}
    \item[]{\sc Amendola, A. and
    Francq, C. } (2009) Concepts and tools for nonlinear time series
    modelling. Handbook of Computational Econometrics, (eds: {\sc D.
    Belsley} and {\sc E. Kontoghiorghes}, Wiley.
    \item[]{\sc Aue, A., H\"{o}rmann S., Horv\`{a}th L.
    and Reimherr, M.} (2009) Break detection in the covariance structure of multivariate time series models.
    \textit{Annals of Statistics} 37, 4046-4087.
    \item[]{\sc
    Andrews, B., Davis, R.A. and Breidt, F.J.} (2006) Maximum
    likelihood estimation for all-pass time series models.
    \textit{Journal of Multivariate Analysis} 97, 1638-1659.


    \item[] {\sc Bai, J. (2000)} Vector autoregressive models with structural changes in
    regression coefficients and in variance-covariance matrices. \textit{Annals of Economics and Finance} 1, 303-339.

    \item[]  {\sc Batbekh, S., Osborn, D.R., Sensier, M., and van Dick, D.} (2007) Is there causality in the mean and volatility of inflation between the US and other G7 countries? Centre for Growth and Business Cycle Research Economic Studies, University of Manchester.


    \item[]  {\sc Berkes, I., Horv\`{a}th, L., and Kokoszka, P.} (2004) Testing for parameter constancy in GARCH(p,q) models. \textit{Statistics and Probability Letters} 70, 263-273.
    \item[]{\sc Blanchard, O., and Simon, J.} (2001). The long and large decline in U.S. output volatility. Brookings Papers on Economic Activity 1, 135-164.

    \item[]{\sc  Bollerslev, T.} (1986) Generalized autoregressive conditional heteroskedasticity. \textit{Journal of Econometrics} 31, 307-327.

    \item[]{\sc Boswijk, H.P., and Zu, Y.} (2007) Testing for cointegration with nonstationary volatility. Working Paper, University of Amsterdam.
    \item[]{\sc Chandler, G., and Polonik, W.} (2012) Mode identification of volatility in time-varying autoregression. \textit{Journal
    of American Statistical Association} 107, 1217-1229.
    \item[]{\sc Dahlhaus, R.} (1997) Fitting time series models to nonstationary processes. \textit{Annals of
    Statistics} 25, 1-37.
    \item[]{\sc Dahlhaus, R., and Subba Rao, S.} (2006) Statistical inference for time-varying ARCH
    processes. \textit{Annals of Statistics} 34, 1075-1114.


    \item[]{\sc Engle, R.F.} (1982) Autoregressive conditional heteroscedasticity with estimates of the variance of United Kingdom inflation. \textit{Econometrica} 50, 987-1007.

    \item[]{\sc  Engle, R.F., and Kroner, K.F.} (1995) Multivariate simultaneous generalized ARCH. \textit{Econometric Theory} 11, 122-150.

    \item[]{\sc Engle, R.F., and Rangel, J.G.} (2008) The spline GARCH model for unconditional volatility and its global macroeconomic causes. \textit{Review of Financial Studies} 21, 1187-1222.

    \item[]{\sc Francq, C., and Gautier, A.} (2004) Estimation of time-varying ARMA models with Markovian changes in regime. \textit{Statistics and Probability Letters} 70, 243-251.
    \item[]{\sc Francq, C., and Ra\"{i}ssi, H.} (2007) Multivariate portmanteau test for
    autoregressive models with uncorrelated but nonindependent errors.
    \textit{Journal of Time Series Analysis} 28, 454-470.
    \item[]{\sc
    Francq, C., Roy, R., and Zako\"{i}an, J-M.} (2005) Diagnostic
    checking in ARMA models with uncorrelated errors. \textit{Journal
    of American Statistical Association} 100, 532-544.

    \item[] {\sc
    Francq, C., and Zako\"{i}an, J-M.} (2010) \textit{GARCH models : structure, statistical inference, and financial applications.}
    Wiley.

    \item[]{\sc Grier, K.B. and Perry, M.J.} (2000) The effects of real and nominal uncertainty on inflation and output growth: some GARCH-M evidence. \textit{Journal of Applied Econometrics} 15, 45-58.
    \item[]{\sc Hafner, C. M. and Linton, O.} (2010) Efficient estimation of a multivariate multiplicative
    volatility model. \textit{Journal of Econometrics} 159, 55-73.


    \item[]{\sc H\"{a}rdle, W. and Marron, J.S.} (1985) Optimal Bandwidth Selection in Nonparametric Regression Function Estimation. \textit{The Annals of Statistics} 13, 1465--1481.

    \item[]{\sc Kim, C.S., and Park, J.Y.} (2010) Cointegrating regressions with time heterogeneity.
    \textit{Econometric Reviews} 29, 397-438.

    \item[] {\sc Kokoszka, P., and Leipus, R.} (2000) Change-point estimation in ARCH models. \textit{Bernoulli} 6, 513-539.
    \item[] {\sc McLeod, A. I., and Li, W. K.} (1983) Diagnostic checking ARMA time series models using squared-residual autocorrelations. \textit{Journal Time Series Analysis} 4, 269-73.
    \item[] {\sc L\"{u}tkepohl, H., Ter\"{a}svirta, T. and Wolters, J.} (1999) Investigating stability and linearity of a German M1 money demand function. \textit{Journal of Applied Econometrics} 14, 511-525.
    \item[]{\sc Mikosch, T., and St\u{a}ric\u{a}, C.} (2004) Nonstationarities in financial time series, the long-
    range dependence, and the IGARCH effects. \textit{Review of Economics and Statistics} 86, 378-390.
    \item[] {\sc Patilea, V., and Ra\"{i}ssi, H.} (2011) Portmanteau tests for stable multivariate autoregressive processes. Working paper   arXiv:1105.3638v2 [stat.ME].
    \item[] {\sc Patilea, V., and Ra\"{i}ssi, H.} (2012) Adaptive estimation of vector autoregressive models with time-varying variance: application to testing linear causality in mean. \textit{Journal of Statistical Planning and Inference} 142, 2891-2912.

    \item[]{\sc Pesaran, H., and Timmerman, A.} (2004) How costly is it to ignore breaks when
    forecasting the direction of a time series. \textit{International Journal of Forecasting} 20,
    411-425.

    \item[]{\sc  Phillips, P.C.B., and Xu, K.L.} (2006)
    Inference in Autoregression under Heteroskedasticity.
    \textit{Journal of Time Series Analysis} 27, 289-308.

    \item[]{\sc Qu, Z., and Perron, P.} (2007) Estimating and testing structural changes in multivariate regressions. \textit{Econometrica} 75, 459-502.
  \item[]{\sc  Ra\"{i}ssi, H.} (2010a) Autocorrelation-based tests for vector error correction models with uncorrelated but nonindependent errors. \textit{Test} 19, 304-324.
    \item[]{\sc  Ra\"{i}ssi, H.} (2010b) Lag length identification for VAR models with non-constant variance. Working paper, IRMAR-INSA.

    \item[]{\sc Sensier, M., and van Dijk, D.} (2004) Testing for volatility changes in U.S. macroeconomic time series. \textit{Review of Economics and Statistics} 86, 833-839.

     \item[]{\sc Sherman, R.P.} (1994)
        Maximal Inequalities for Degenerate U-Processes with Applications to Optimization Estimators. \textit{The Annals of Statistics} 22,  439-459.

    \item[]{\sc Spokoiny, V.} (2009) Multiscale local change point detection with applications to value-at-risk. \textit{The Annals of Statistics} 37, 1405-1436.
    \item[]{\sc St\u{a}ric\u{a}, C.} (2003) Is GARCH(1,1) as good a model as the Nobel
    prize accolades would imply? Working paper, \\http://129.3.20.41/eps/em/papers/0411/0411015.pdf.
    \item[]{\sc St\u{a}ric\u{a}, C., and Granger, C.} (2005) Nonstationarities in stock returns. \textit{Review of Economics and Statistics} 87, 503-522.

    \item[]{\sc van der Vaart, A.W., and Wellner, J.A.} (1996). \emph{Weak convergence and empirical processes.}
Springer Series in Statistics. Springer-Verlag, New-York.

    \item[]{\sc Warnock, M.V.C., and Warnock, F.E.} (2000) The declining volatility of U.S. employment: was Arthur Burns right? Board of Governors of the Federal Reserve System, International finance discussion papers, 677.
    \item[]{\sc Xu, K.L., and Phillips, P.C.B.} (2008)
    Adaptive estimation of autoregressive models with time-varying
    variances. \textit{Journal of Econometrics} 142, 265-280.
    \end{description}

    \newpage

    \section{Appendix}
    \label{SApp}

    \subsection{Proof of Proposition \ref{propostu}}
    The proof being quite straightforward, we only outline the arguments. By the Mean Value Theorem
    \begin{equation}\label{blues}
    u_t^2(\hat{\theta})=u_t^2+2u_t\underline{x}_{t-1}'(\hat{\theta}-\theta_0)
    -2(\theta^*-\theta_0)'\underline{x}_{t-1}\underline{x}_{t-1}'(\hat{\theta}-\theta_0)
    \end{equation}
    where $\theta^*$ is between $\theta_0$ and $\hat{\theta}.$ Recall that $u_t=u_t(\theta_0)$ and  $u_t(\hat{\theta})$ corresponds to the GLS residuals. By the Taylor expansion of  $\mathcal{S}(\hat{\theta}),$ since the $u_t$'s are independent, $\sqrt{n}(\hat{\theta}-\theta_0)=O_p(1),$ and using similar arguments to those in Lemma 1 and 2 of Phillips and Xu (2006),  it can be shown that $\mathcal{S}(\hat{\theta})=\mathcal{S}(\theta_0)+o_p(1)$. Since
    $\left\{\left(u_t^2(\theta_0)/ h_t^2
    -1\right)U_{t-1}(\theta_0)\right\}$ is a martingale difference
    sequence, it follows from the  Lindeberg CLT that
    $\mathcal{S}(\theta_0)$ is asymptotically normally distributed.
    Noting that the process $(\epsilon_t^2-1)$ is independent with mean
    zero and finite variance we readily obtain the form of
    the asymptotic covariance matrix
    \begin{eqnarray*}
    \mbox{Var}_{as}(S(\theta_0))&=&\lim_{n\to\infty}\frac{1}{4n}
    \sum_{t=1}^n E\left\{\left(\frac{u_t^2}{h_t^2}-1\right)^2U_{t-1}(\theta_0)U_{t-1}(\theta_0)'\right\}\\
    &=&\frac{\mbox{Var}(\epsilon^2)}{4}\lim_{n\to\infty}\frac{1}{n}
    \sum_{t=1}^n \left( E\left\{\frac{u_{t-i}^2u_{t-j}^2}{h_t^4}\right\}\right)_{1\leq i,j\leq m}.
    \end{eqnarray*}
    Considering again  arguments like in the proof of Lemmas 1 and 2 in Phillips and Xu (2006) deduce that the limit exists and is equal to the  positive definite matrix $\Sigma$.


    \subsection{Proof of Proposition \ref{proposthree}}
    The proofs of the asymptotic equivalences $Q_{ALS}=Q_{GLS}+o_p(1)$ and $Q_{ALS}^*=Q_{GLS}^*+o_p(1)$
    are direct consequences of the following uniform rates
    \begin{equation}\label{err1}
    \sup_{b\in\mathfrak{B}_n}\left|\frac{1}{n}\sum_{t=1}^n \{u_t^s( \theta_n)  \hat h_t ^{-s} - u_t^s   h_t ^{-s}\}\right| = o_p (1), \quad s=2,4,8,
    \end{equation}

    \begin{equation}\label{err1e2}
    \sup_{1\leq j\leq m}\sup_{b\in\mathfrak{B}_n}\!\left|\frac{1}{\sqrt{n}}\sum_{t=j+1}^n \{(u_{t}^2(\theta_n)\!  -\!\hat h_{t} ^{2})(u_{t-j}^2(\theta_n  \!-\!\hat h_{t-j} ^{2}) - ( u_{t}^2 \!- \! h_{t} ^{2})( u_{t-j}^2 \!-\!  h_{t-j} ^{2})\}\right| \!=\! o_p (1),
    \end{equation}

    \begin{equation}\label{err4}
    \sup_{1\leq j\leq m}\sup_{b\in\mathfrak{B}_n}\left|\frac{1}{\sqrt{n}}\sum_{t=j+1}^n \{(u_t^2(\theta_n)\hat h_t ^{-2} -1) u_{t-j}^2(\theta_n) \hat h_t ^{-2} \!-\! (u_t^2 h_t ^{-2} -1)  u_{t-j}^2 h_t ^{-2}\}\right| = o_p (1),
    \end{equation}
where $\theta_n -\theta_0 = O_p(1/\sqrt{n})$. We will focus on equations (\ref{err1e2}) and (\ref{err4}), the arguments for proving (\ref{err1}) being similar and much shorter.
    For deriving these rates we will use the following lemma. Below $[a]$ denotes the integer part of the real number $a$ (that is, the largest integer smaller than $a$).

    \begin{lem}\label{bias_h2}
    Suppose that the assumptions of Proposition \ref{proposthree} hold true. Let $$G(r) = g^2(r-)\int_{-\infty}^0 K(z) dz + g^2(r+)\int_{0}^\infty K(z) dz ,\quad r\in[0,1],$$    with $g(r-)=\lim_{\overline{r}\uparrow r} g(\overline{r})$ and $g(r+)=\lim_{\overline{r}\downarrow r} g(\overline{r})$ for $r\in(0,1),$   $g(0-)=g(1+)=0.$
    \begin{enumerate}

    \item[(a)] Let $k=0$ or $k=1$ and
    \begin{eqnarray*}
    s_n(t/n;b) &= &\frac{1}{nb}\sum_{i=1}^n \left|\frac{t-i}{nb}\right|^k K((t-i)/nb),\;\;\; 1\leq t\leq n, \\
      s(r,b) &= &\int _{(r-1)/b}^{r/b} |z|^k K(z) dz,\;\;\; r\in(0,1].
    \end{eqnarray*}
There exists a constant $C>0$ independent of $n$ and $b_n$ such that
\begin{equation}\label{zert}
\sup_{b\in\mathfrak{B}_n}\sup_{1\leq t\leq n}|s_n(t/n;b) - s(t/n,b)| \leq \frac{C}{nb_n}.
\end{equation}

    \item[(b)]
    $$
     \sup_{b\in\mathfrak{B}_n}\sup_{r\in(0,1]} \left| \{\hat h_{[nr]} ^{2} -h_{[nr]} ^{2}  \} -D(r)- \Lambda_n(r) - \Gamma_n(r)
    \right| = o_p(1/\sqrt{n}),
    $$
    where $D(r)
     = G(r)-g^2(r)$
    $$
    \Lambda_n(r) = \sum_{i=1}^n w_{[nr]i} (\epsilon_i^2 -1)h_i^2 \quad \text{ and }\quad\Gamma_n(r) = \sum_{i=1}^n w_{[nr]i} h_i^2 - G(r).
    $$
   Moreover, $D(r)=0$ for all but  a finite number of values $r\in(0,1],$
    $$
    \sup_{b\in\mathfrak{B}_n}\sup_{r\in(0,1]} \left| \Lambda_n(r) \right|= O_p\left(  \frac{\sqrt{\ln (1/ b_n)}}{\sqrt{nb_n}} \right) \quad \text{ and } \quad \sup_{b\in\mathfrak{B}_n}\sup_{r\in(0,1]} \left| \Gamma_n(r) \right| \leq C b_n,
    $$
    for some constant $C>0$ independent of $n$ and $b_n.$ In particular
    $$
    \sup_{b\in\mathfrak{B}_n}\sup_{r\in(0,1]}  \left| \hat h_{[nr]} ^{2} - G(r)\right| = o_p(1).
    $$
    \end{enumerate}
    \end{lem}

\bigskip

Now, let us justify (\ref{err1e2}).  Let $\Delta_t$ be a short notation for $ \hat h_t^2 -  h_t^2$ and fix $1\leq j \leq m.$
    Using the decomposition (\ref{blues}) and the moment conditions on the process $(\epsilon_t),$ $\forall t$
    \begin{multline*}
    (u_t^2(\theta_n)-
    \hat h_t^2)(u_{t-j}^2(\theta_n)-\hat h^2_{t-j})= (u_t^2-
    h_t^2)(u_{t-j}^2- h^2_{t-j})\\
    + \Delta_t\Delta_{t-j} + (u_t^2-
    h_t^2) \Delta_{t-j}  + (u_{t-j}^2- h^2_{t-j}) \Delta_t\\
    +2\{(u_t^2-
    h_t^2)u_{t-j}\underline{x}_{t-j-1}'  + (u_{t-j}^2- h^2_{t-j}) u_t \underline{x}_{t-1}' \}(\theta_n - \theta_0)\\
    + \|\theta_n - \theta_0\|^2 R_t
    \end{multline*}
    with $\sup_{t}|R_t|=O_p(1).$ By simple calculations of the mean and the variance, we deduce
    $$
    \frac{1}{n}\sum_{t} \{(u_t^2-
    h_t^2)u_{t-j}\underline{x}_{t-j-1}'  + (u_{t-j}^2- h^2_{t-j}) u_t \underline{x}_{t-1}' \} = O_p (1/\sqrt{n}).
    $$
Next, as a direct consequence of Lemma \ref{bias_h2} and the conditions on $b_n$,
$$
\frac{1}{n}\sum_{t}\Delta_t\Delta_{t-j} = O_p\left( \frac{\ln(1/b_n)}{nb_n} +
\frac{\sqrt{b_n\ln(1/b_n)}}{\sqrt{n}}\right)+ O(b_n^2)=o_p (1/\sqrt{n}),
$$
uniformly for $b\in\mathfrak{B_n}.$ Finally, let us decompose
\begin{eqnarray*}
\frac{1}{n}\sum_{t} (u^2_t - h_t^2)\Delta_{t-j}&=& \frac{1}{n}\sum_{t,i}w_{ti}(\epsilon^2_t -1)(\epsilon^2_i -1)h_t^2h_i^2 \\
&&+ \frac{1}{n} \sum_{t} (u^2_t - h_t^2)\{\Gamma_n((t-j)/n)+ D((t-j)/n)\}\\
&\stackrel{def}{=}& C_{1n}+ C_{2n},
\end{eqnarray*}
with $\Gamma_n(\cdot)$ and $D(\cdot)$ defined in Lemma \ref{bias_h2}. Take absolute values, use the uniform bound for $\Gamma_n(\cdot)$, the classical Law of large Numbers and the fact that $D(r)\equiv 0$  except a finite set of values $r\in(0,1].$ Deduce that
$\sup_{b\in\mathfrak{B}_n}|C_{2n} | = O_p(b_n^2)=o_p (1/\sqrt{n}).$ Next, decompose
\begin{eqnarray*}
C_{1n} &=& \frac{1}{n^2b}\sum_{t,i}K\!\left(\frac{t-i}{nb}\right)\frac{1}{s(t/n;b)}(\epsilon^2_t -1)(\epsilon^2_i -1)h_t^2h_i^2 \\
&&+ \frac{1}{n^2b}\sum_{t,i} K\!\left(\frac{t-i}{nb}\right) \frac{s(t/n;b) -s_n(t/n;b) }{s(t/n;b)s_n(t/n;b)} (\epsilon^2_t -1)(\epsilon^2_i -1)h_t^2h_i^2\\
&\stackrel{def}{=}& C_{11n}+C_{12n}.
\end{eqnarray*}
Take absolute values and expectation and use the bound (\ref{zert}) three times to deduce
$$
\sup_{b\in\mathfrak{B}_n}\sup_{1\leq t\leq n}|C_{12n}| \leq \frac{C}{nb_n}=o_p (1/\sqrt{n}).
$$
Finally, note that $bC_{11n}$ is a degenerate $U-$process of order 2 indexed by a VC family of functions. Again, for justifying the VC see the proof of Proposition 4.1 in Patilea and Ra\"{i}ssi (2012a). Use the uniform rate of such a process, see for instance Sherman (1994), and deduce that
$$
\sup_{b\in\mathfrak{B}_n}\sup_{1\leq t\leq n}|C_{12n}|=O_p (1/nb)=o_p (1/\sqrt{n}).
$$
Gathering facts deduce the equivalence (\ref{err1e2}).

Now, let us justify (\ref{err4}). Use the decomposition (\ref{blues}), some long but elementary algebra and arguments that have been already used several times above to write
$$
\left(\frac{u_t^2(\theta_n)}{\hat{h}_t^2}-1\right)
\frac{u_{t-j}^2(\theta_n)}{\hat{h}_t^2} -     \left(\frac{u_t^2}{h_t^2}-1\right)
\frac{u_{t-j}^2}{h_t^2} =  R_{1t,j} + \|\theta_n - \theta_0\| R_{2t,j} + O_p(b_n^2),
$$
with
$$
R_{1t,j}= \frac{(h_t^2 - \hat h_t^2) (\epsilon_t^2 -1) u_{t-j}^2}{h_t^4}
$$
and
$\sup_{b\in\mathfrak{B}_n}\sup_{1\leq t\leq n} \left|R_{2t,j}\right| = o_p(1).$
By Lemma \ref{bias_h2}, $\sup_{b\in\mathfrak{B}_n}\sup_{1\leq t\leq n} \left|R_{1t,j}\right| = o_p(1/\sqrt{n}).$ Deduce that
\begin{eqnarray*}
\mathcal{\widetilde{S}}_j(\theta)&= & \frac{1}{2\sqrt{n}} \sum_{t=1}^n\left(\frac{u_t^2}{h_t^2}-1\right)\frac{u_{t-j}^2}{h_t^2}+ \frac{1}{2\sqrt{n}} \sum_{t=1}^n R_{1t,j} + o_p(1)\\
 &= &\mathcal{S}_j(\theta_0) + o_p(1),
\end{eqnarray*}
from which the equivalence (\ref{err4}) follows.
    \quad$\square$\\

\bigskip

    \noindent{\bf Proof of Lemma \ref{bias_h2}}

(a) Let $\widetilde K(x)=|x|^k K(x)$ with $k=0$ or $k=1.$
 Then, for any  $1\leq t\leq n$
    \begin{eqnarray}\label{eq9}
\left| s_n (t/n;b) -  s(t/n;b)\right|
&\leq & \int_{\frac{t-n}{nb}}^{\frac{t}{nb}}
 \left|  \widetilde K\left(
\frac{[nzb]}{nb} \right) -\widetilde K\left( z \right) \right| dz\\
&\leq & \frac{C_1}{nb_n} \int_{\frac{t-n}{nb}}^{\frac{t}{nb}} \left| \widetilde K^\prime \left( z \right) \right|dz\notag \\
&\leq& \frac{C_2}{nb_n},\notag
\end{eqnarray}
for some constants $C_1, C_2>0$ independent of $t,$  $n$ and $b_n$.

(b) By the decomposition (\ref{blues}), for any $1\leq t\leq n$
\begin{eqnarray*}
\hat h_t^2 - h_t^2 - \Lambda_n(t/n) - \Gamma_n(t/n) - D(t/n)
  &=& 2\sum_{i=1}^n w_{ti} u_i \underline{x}_{i-1}'(\theta_n-\theta_0) \\
  &&+
 \|\theta_n - \theta_0 \|^2 O_p(1)\sum_{i=1}^n w_{ti}\|\underline{x}_{i-1} \|^2\\
 &\stackrel{def}{=}& 2\delta_{1t}'(\tilde{\theta}\!-\!\theta_0) + \| \theta_n \!-\! \theta_0 \|^2 O_p(1)\delta_{2t},
\end{eqnarray*}
where the $O_p(1)$ term is independent of $n$ and $b$. Next, note that
$\delta_{1t}$ is a sum of centered zero-covariance  random vectors. For a given sequence of bandwidths $b$ the variance of $\delta_{1t}$ tends to zero, uniformly with respect to $t$. To derive the uniform rate for $\delta_{1[nr]}$
one could first replace $(1/nb)\sum_{i} K(([nr]-i)/nb)$ in the denominator  of $w_{[nr]i}$ by $\int _{(r-1)/b}^{r/b}  K(z) dz.$ Let  $\bar \delta_{1[nr]}$ be the quantity obtained after this change of denominators.
Proceeding like in equation (\ref{geza}) below, it is easy to see that   $ \delta_{1[nr]} -\bar \delta_{1[nr]} =o_p(1)$ uniformly with respect to to $1 < r \leq 1$ and $b\in\mathfrak{B}_n.$ To get the uniform rate of $ \bar \delta_{1[nr]}$ one can use the uniform Law of Large Numbers  for Vapnik-Chervonenkis (VC) classes of functions (see for instance van der Vaart and Wellner, 1996). The VC property could be established following the lines of the proof of Proposition 4.1 in Patilea and Ra\"{i}ssi (2012a). Finally, by the definition of the weights $w_{ti}$ and the moment conditions on the $\epsilon_t$'s, we have $ \delta_{2t}\leq \max_i \|\underline{x}_{i-1} \|^{2} = o_p(\sqrt{n}),$  uniformly with respect to $b$. Since $g(\cdot)$ is continuous except a finite set of points in $(0,1],$ deduce  $D(r)=0$ for all but  a finite number of values $r\in(0,1]$.

To  derive the uniform bound for $\Gamma_n(\cdot)$ let us write for $r\in[0,1],$
$$
\Gamma_n(r)  =  \left[\sum_{i <[nr]} w_{[nr]i} h_i^2 -  c_Kg^2(r-)\right]
+ \left[\sum_{i  > [nr]} w_{[nr]i} h_i^2 -  (1-c_K) g^2(r+)\right],
$$
where $c_K=\int_{-\infty}^0 K(z) dz$ and by definition $\sum_{i <0} \dots = \sum_{i >1} \dots  = 0.$ Next, if $r>0$ we write
$$
\left|\sum_{i <[nr]} w_{[nr]i} h_i^2 -  c_K g^2(r-)  \right| \leq
\sum_{i <[nr]} \!\!w_{[nr]i} \left| h_i^2 -  g^2(r-)  \right| +  g^2(r-)
\left| \sum_{i <[nr]} \!\!w_{[nr]i} -  c_K \right|.
$$
On  the other hand, by the arguments used for equation (\ref{eq9}), the last absolute value is of uniform order $1/nb_n.$ Using the piecewise Lipschitz property  and again inequalities like in equation (\ref{eq9}), there exists $0 \leq r_1 < r$ and some constants $C_1,C_2,C_3>0$ such that
\begin{eqnarray*}
\sum_{i <[nr]}\!\! w_{[nr]i} \!\left| h_i^2 \!-  g^2(r-)  \right|\!\! \!&=&\!\!
\sum_{[nr_1]<i <[nr]} w_{[nr]i}\left| h_i^2 -  g^2([nr]/n)  \right| \\&+&\!\!\!
| g^2([nr]/n) \! - \! g^2(r-)| \!\!\sum_{i <[nr]} w_{[nr]i}  \!+
2\sup_{0<r\leq 1} g^2(r) \!\!\!\sum_{i \leq[nr_1]} w_{[nr]i} \\
&\leq & \!\!\! C_1b_n \sum_{[nr_1]<i <[nr]} w_{[nr]i}\left|\frac{i -  [nr]}{nb}\right| + \frac{C_2}{nb_n}\\
&\leq & \!\!\! C_3\left[ b_n \int_{r_1}^r |z|K(z) dz + 1/nb_n \right].
\end{eqnarray*}
Similarly, if $r<1$ there exists $r<r_2\leq 1$ and some constant $C_4$ such that
$$
\left|\sum_{i  > [nr]} w_{[nr]i} h_i^2 -  (1-c_K) g^2(r+)\right|\leq C_4\left[ b_n \int^{r_2}_r |z|K(z) dz + 1/nb_n \right].
$$
Deduce the uniform rate for $\Gamma_n(\cdot).$

Next, for a given sequence of bandwidths and a fixed $r$, up to the logarithm factor, the rate of $\Lambda_n(r)$ is a standard asymptotic result in nonparametric regression. To derive the uniform rate for $\Lambda_n(\cdot),$  consider $s_n(r;b)$ defined with $k=0$ and the corresponding $s(r;b)$ that is constant equal to 1 for $r\in(0,1)$, $s(0+;b)=c_K$ and $s(1-;b)=1-c_K$.
Use the simple identity
$x^{-1}= y^{-1} + x^{-1}(y-x)y^{-1},$ $x,y\neq 0$,  and write
\begin{eqnarray}\label{geza}
\Lambda_n(t/n) \!\!&=& \frac{1}{s(t/n;b)}\frac{1}{nb}\sum_{i=1}^n K\!\left(\frac{t-i}{nb}\right) (\epsilon_i^2 -1)h_i^2\\
&&\!\!\!\!+\; \frac{1}{s(t/n;b)s_n(t/n;b)}\;\frac{1}{nb}\sum_{i=1}^n K\!\left(\frac{t-i}{nb}\right) \{s(t/n;b) -s_n(t/n;b) \}(\epsilon_i^2 -1)h_i^2\notag\\
&\stackrel{def}{=}& \bar \Lambda_n( t/n ) + R_n(t/n)\notag.
\end{eqnarray}
Take absolute values and expectation and use the bound (\ref{zert}) three times and deduce
$$\sup_{b\in\mathfrak{B}_n}\sup_{1\leq t\leq n}|R_n(t/n)|=O_p(1/nb_n)=o_p(1/\sqrt{n}).$$
The uniform rate of $\bar \Lambda_n(\cdot)$ could be obtained from Theorem 2.14.16 of van der Vaart and Wellner (1996). \quad$\square$\\

\newpage

\section{Appendix B: Tables and Figures}

\begin{table}[hh]\!\!\!\!\!\!\!\!\!\!
\begin{center}
\caption{\small{Empirical size (in \%) of the tests for second order dynamics. The innovations are
heteroscedastic.}}
\begin{tabular}{c|c|c|c||c|c|c||c|c|c|}\cline{2-10}
 & \multicolumn{3}{|c||}{$m=1$}&\multicolumn{3}{|c||}{$m=3$}&\multicolumn{3}{c|}{$m=6$}\\
\hline
  \multicolumn{1}{|c|}{n} & 100 & 200 & 500  &100 & 200 & 500&100 & 200 & 500 \\
  \hline
  \multicolumn{1}{|c|}{$LM_{S}$}  & 10.4 & 14.8 & 26.0 & 16.6 & 23.6 & 40.0 & 17.1 & 26.2&47.2\\
  \hline
  \multicolumn{1}{|c|}{$LM_{GLS}$}  & 4.8 & 5.2 & 5.6 & 4.8 & 5.5 & 5.6 & 5.4 & 5.7& 4.8\\
  \hline
  \multicolumn{1}{|c|}{$LM_{cv,ALS}$}  & 0.3 & 1.1 & 0.8 & 0.9 & 1.3 & 1.2 & 1.3 &1.2 & 2.4\\
  \multicolumn{1}{|c|}{$LM_{cv,ALS}^{mc}$}    & 0.2 & 0.9 & 3.2 & 0.0 & 0.1 & 0.4 & 0.0 & 0.0&0.0\\
  \hline
  \multicolumn{1}{|c|}{$LM_{sm,ALS}$}    & 1.2 & 1.9 & 1.6 & 1.7 & 2.4 & 2.6 & 1.8 & 2.2&2.4\\
  \multicolumn{1}{|c|}{$LM_{sm,ALS}^b$}    & 4.3 & 4.9 & 5.5 & 2.3 & 4.4 & 4.8 & 2.3 & 2.7&5.0\\
  \multicolumn{1}{|c|}{$LM_{sm,ALS}^{mc}$}    & 0.3 & 0.7 & 1.5 & 0.0 & 0.0 & 0.3 & 0.0 & 0.0&0.1\\
  \hline
  \multicolumn{1}{|c|}{$LM_{la,ALS}$}    &  2.8& 3.1 & 2.6 & 3.6 & 4.2 & 4.1 & 3.2 & 4.2&4.3\\
  \multicolumn{1}{|c|}{$LM_{la,ALS}^b$}    & 5.5 & 5.8 & 6.0 & 4.0 & 5.7 & 5.7 & 3.1 & 4.0&6.7\\
  \multicolumn{1}{|c|}{$LM_{la,ALS}^{mc}$}    & 0.3 & 0.7 & 1.4 & 0.0 & 0.3 & 0.4 & 0.0 & 0.0&0.2\\
  \hline\hline
  \multicolumn{1}{|c|}{$LB_{S}$}    & 10.8 & 15.0 & 26.4 & 16.7 & 25.3 & 44.0 & 19.2 & 30.0&49.2\\
  \hline
    \multicolumn{1}{|c|}{$LB_{GLS}$}    & 6.2 & 6.4 &  5.3& 9.5 & 8.4 &  6.2& 10.3 &10.2 &7.3\\
  \hline
  \multicolumn{1}{|c|}{$LB_{cv,ALS}$}    & 5.8 & 6.8 & 6.0 & 9.1 & 7.9 & 7.6 & 12.4 &10.2 &7.2\\
  \multicolumn{1}{|c|}{$LB_{cv,ALS}^{mc}$}    & 7.8 & 7.9 & 8.4 & 5.2 & 4.4 & 5.6 & 6.1 & 4.7&5.6\\
  \hline
  \multicolumn{1}{|c|}{$LB_{sm,ALS}$}    & 6.0 & 7.2 & 5.6 & 8.0 & 7.9 & 7.3 & 9.8 & 10.2&8.8\\
  \multicolumn{1}{|c|}{$LB_{sm,ALS}^b$}    & 6.6 & 6.2 & 6.5 & 6.6 & 6.7 & 7.5 & 7.3 & 7.2&6.5\\
  \multicolumn{1}{|c|}{$LB_{sm,ALS}^{mc}$}    & 7.4 & 6.9 & 8.6 & 4.7 & 4.0 & 4.4 & 5.2 & 4.2&4.5\\
  \hline
  \multicolumn{1}{|c|}{$LB_{la,ALS}$}    & 6.8 & 7.3 & 6.2 & 8.7 & 9.2 & 8.8 & 10.2 & 11.6&10.8\\
  \multicolumn{1}{|c|}{$LB_{la,ALS}^b$}    & 7.6 & 8.2 & 9.1 & 10.1 & 9.0 & 10.7 & 10.6 & 11.2&11.9\\
  \multicolumn{1}{|c|}{$LB_{la,ALS}^{mc}$}    & 5.0 & 6.5 & 6.2 & 3.5 & 3.6 & 3.9 & 3.7 & 3.6&4.0\\
  \hline
\end{tabular}
\label{tab2}
\end{center}
\end{table}

\begin{table}[hh]\!\!\!\!\!\!\!\!\!\!
\begin{center}
\caption{\small{Empirical size (in \%) of the tests for second order dynamics. The innovations are
homoscedastic.}}

\medskip

\begin{tabular}{c|c|c|c||c|c|c||c|c|c|}\cline{2-10}
 & \multicolumn{3}{|c||}{$m=1$}&\multicolumn{3}{|c||}{$m=3$}&\multicolumn{3}{c|}{$m=6$}\\
\hline
  \multicolumn{1}{|c|}{n} & 100 & 200 & 500  &100 & 200 & 500&100 & 200 & 500 \\
  \hline
  \multicolumn{1}{|c|}{$LM_{S}$}  & 3.5 & 4.5 & 4.9 & 3.9 & 4.3 & 5.2 & 4.1 & 4.0&4.5\\
  \hline
  \multicolumn{1}{|c|}{$LM_{cv,ALS}$}  & 0.3 & 1.5 & 1.8 & 1.2 & 1.7 & 3.3 & 1.7 & 2.3&2.2\\
  \multicolumn{1}{|c|}{$LM_{cv,ALS}^{mc}$}    & 0.3 &  0.9& 1.6 & 0.0 &  0.1& 0.4 & 0.0 & 0.0&0.1\\
  \hline
  \multicolumn{1}{|c|}{$LM_{sm,ALS}$}    & 0.9 & 1.7 & 1.7 & 1.2 & 1.7 & 3.0 & 1.6 & 1.8&2.0\\
  \multicolumn{1}{|c|}{$LM_{sm,ALS}^b$}    & 4.0 & 4.5 & 5.6 & 2.5 & 3.1 & 5.0 & 2.3 & 2.2&5.1\\
  \multicolumn{1}{|c|}{$LM_{sm,ALS}^{mc}$}    & 0.3 & 0.7 & 1.4 & 0.0 & 0.0 & 0.4 & 0.0 & 0.0&0.1\\
  \hline
  \multicolumn{1}{|c|}{$LM_{la,ALS}$}    & 0.7 & 2.0 & 1.8 & 1.7 & 2.1 & 3.4 & 2.3 & 2.0&2.5\\
  \multicolumn{1}{|c|}{$LM_{la,ALS}^b$}    & 3.9 & 4.4 & 5.9 & 2.9 & 3.8 & 5.3 & 2.4 & 2.8&5.7\\
  \multicolumn{1}{|c|}{$LM_{la,ALS}^{mc}$}    & 0.3 & 0.8 & 1.7 & 0.0 & 0.0 & 0.4 & 0.0 &0.0 &0.1\\
  \hline\hline
  \multicolumn{1}{|c|}{$LB_{S}$}    & 3.5 & 4.8 & 4.9 & 4.4 & 4.5 & 5.3 & 5.1 & 4.1&4.8\\
  \hline
  \multicolumn{1}{|c|}{$LB_{cv,ALS}$}    & 3.6 & 5.3 & 4.6 & 5.3 & 4.5 & 5.7 & 6.1 & 4.7&5.5\\
  \multicolumn{1}{|c|}{$LB_{cv,ALS}^{mc}$}    & 6.6 &  6.9& 7.1 & 5.0 & 4.6 & 5.4 & 4.6 & 4.2&5.4\\
  \hline
  \multicolumn{1}{|c|}{$LB_{sm,ALS}$}    & 4.9 & 6.0 & 5.7 & 6.2 & 5.7 & 6.0 & 7.1 & 5.5&5.9\\
  \multicolumn{1}{|c|}{$LB_{sm,ALS}^b$}    & 4.3 & 4.8 & 5.2 & 4.2 & 4.2 & 4.9 & 4.2 & 3.8&5.8\\
  \multicolumn{1}{|c|}{$LB_{sm,ALS}^{mc}$}    & 8.1 & 7.9 & 8.3 & 6.9 & 5.2 & 5.7 & 6.2 & 4.6&5.5\\
  \hline
  \multicolumn{1}{|c|}{$LB_{la,ALS}$}    & 4.5 & 5.1 & 5.1 & 5.1 & 5.0 & 5.7 & 5.7 & 4.7&5.5\\
  \multicolumn{1}{|c|}{$LB_{la,ALS}^b$}    & 4.7 & 5.0 & 5.3 & 4.5 & 4.4 & 5.5 & 5.0 & 3.9&5.9\\
  \multicolumn{1}{|c|}{$LB_{la,ALS}^{mc}$}    & 6.6 & 6.6 & 7.3 & 5.5 & 4.6 & 5.1 & 4.6 & 3.8&4.8\\
  \hline
\end{tabular}
\label{tab1}
\end{center}
\end{table}

\begin{table}[hh]\!\!\!\!\!\!\!\!\!\!
\begin{center}
\caption{\small{The $p$-values (in \%) of the tests for testing
the presence of second order dynamics in the residuals
of the M1 and PPI data.
}}

\medskip

\begin{tabular}{c|c|c|c||c|c|c|}\cline{2-7}
&\multicolumn{3}{|c||}{M1}&\multicolumn{3}{|c|}{PPI}\\
\cline{2-7}
 & \multicolumn{1}{|c|}{$m=1$}&\multicolumn{1}{|c|}{$m=3$}&\multicolumn{1}{c||}{$m=5$}
 & \multicolumn{1}{|c|}{$m=1$}&\multicolumn{1}{|c|}{$m=3$}&\multicolumn{1}{c|}{$m=5$}\\
  \hline
  \multicolumn{1}{|c|}{$LM_{S}$}  & {\bf 0.0} &{\bf 0.0}& {\bf 0.0}    & {\bf 0.0}& {\bf 0.0}& {\bf 0.0}\\
  \hline\hline
  \multicolumn{1}{|c|}{$LB_{S}$}  & {\bf 0.0} &{\bf 0.0}& {\bf 0.0}    & {\bf 0.0}& {\bf 0.0}& {\bf 0.0}\\
  \multicolumn{1}{|c|}{$LB_{cv,ALS}^{mc}$}  & 76.2 &28.6 &25.0     & 36.6& 52.6& 34.0\\
  \hline
\end{tabular}
\label{pvaluestsod1}
\end{center}
\end{table}

\begin{table}[hh]\!\!\!\!\!\!\!\!\!\!
\begin{center}
\caption{\small{The $p$-values (in \%) of the tests for testing
the presence of second order dynamics in the residuals
of the FDI and CPI data.
}}

\medskip

\begin{tabular}{c|c|c|c|}\cline{2-4}
&\multicolumn{3}{|c|}{CPI}\\
\cline{2-4}
 & \multicolumn{1}{|c|}{$m=1$}&\multicolumn{1}{|c|}{$m=3$}&\multicolumn{1}{c|}{$m=5$}\\
  \hline
  \multicolumn{1}{|c|}{$LM_{S}$}   & {\bf 0.0}& {\bf 0.0}& {\bf 0.0}\\
  \multicolumn{1}{|c|}{$LM_{la,ALS}^{b}$} & {\bf 3.5}& 7.6& 16.2\\
  \hline\hline
  \multicolumn{1}{|c|}{$LB_{S}$}  & {\bf 0.0}& {\bf 0.0}& {\bf 0.0}\\
  \multicolumn{1}{|c|}{$LB_{cv,ALS}^{mc}$}  & 49.0& 84.0& 66.2\\
  \hline
  \multicolumn{1}{|c|}{$LB_{sm,ALS}^{mc}$}  & 17.4& 99.2& 59.4\\
  \hline
\end{tabular}
\label{pvaluestsod2}
\end{center}
\end{table}

\clearpage

\end{document}